\documentclass[preprint,amsmath,amssymb,aps,prl]{revtex4-1}

\usepackage{graphicx}
\usepackage{color}
\usepackage{dcolumn}
\usepackage{bm}
\usepackage{hyperref}
\usepackage[mathlines]{lineno}

\begin{document} 

\title{Development of ion recoil energy distributions in the Coulomb explosion of argon clusters resolved by charge-state selective ion energy spectroscopy}

\author{D. Komar$^{1}$, L. Kazak$^{1}$, K.-H. Meiwes-Broer$^{1,2}$, and J. Tiggesb\"aumker$^{1,2}$}

\address{$^1$University of Rostock, Albert-Einstein-Strasse 23, 18059
  Rostock, Germany\\$^2$Department ''Life, Light, and Matter'', Universit\"at Rostock, 18059 Rostock, Germany }

\begin{abstract}
The laser intensity dependence of the recoil energies from the Coulomb explosion of small argon clusters has been investigated by resolving the contributions of the individual charge states to the ion recoil energy spectra. Between $10^{14}$ and $10^{15}$\,W/cm$^2$ the high-energy tail of the ion energy spectra changes its shape and develops into the well-known \emph{knee} feature, which results from the cluster size distribution, laser focal averaging, and ionization saturation. Resolving the contributions of the different charge states to the recoil energies, the experimental data reveal that the basic assumption of an exploding homogeneously charged sphere cannot be maintained in general. In fact, the energy spectra of the high-$q$ show distinct gaps in the yields at low kinetic energies, which hints at more complex radial ion charge distributions developing during the laser pulse impact.

\end{abstract}

\maketitle

\section{Introduction}
\label{sec:intro}

Recent developments in ultrafast laser technology provide opportunities to study light-matter interactions in the regime, where the energy absorption depends on the field strength rather than on the photon energy. Laser-produced plasmas from optically ionized solid density matter have been widely explored over the past years as potential sources for energetic particles~\cite{EsaTPS96} and short wavelength radiation~\cite{KmePRL92}. Small particles can be utilized as a replacement for solid targets as they provide debris-free conditions in the experiment as well as yet retain solid-like properties. Measurements on clusters are appealing since they act as nm-sized model systems to study base issues of strongly non-linear many-body interactions with light~\cite{KraPR02,SaaJPB06,FenRMP10}, such as the impact of collective effects on the charging dynamics~\cite{DitPRA96,SaaPRL03,FenPRL07a}. Emission of highly energetic ions in various charge states~\cite{SnyPRL96,LezPRL98} and fast electrons~\cite{SprPRA03} was observed from laser heated clusters. Moreover, the control of fast electron emission on the attosecond timescale through modification of the collective oscillations of the nanoplasma by two-color laser field has been reported~\cite{PasNCom17}.

The ionization dynamics of rare gas clusters are in the  focus of experimental studies~\cite{LezPRL98,HirPRA04,RajNPhys13,RupPRL16,KumPRX18,ZhaJPCL20,SchPRL18,StrAPL18}. To date, only a few experiments aimed to resolve the underlying charge-state distributions upon exposure to strong optical laser fields~\cite{LezPRL98,HirPRA04,RajNPhys13}. In these measurements, however, laser intensities above $10^{16}$ W/cm$^2$ have been utilized, which is about two orders of magnitude higher than the atomic barrier suppression intensity threshold $I_{BSI}$~\cite{AugPRL89}. $I_{BSI}$ can be taken as a measure for the onset of extreme cluster charging~\cite{TruEPJD11}. The threshold regime is appealing since one can expect to resolve, e.g., the transition from a partially to a fully ionized plasma.    

Experimental data of the Coulomb explosion of clusters were analyzed by Islam et al.~\cite{IslPRA06}, to explain the recoil energy spectra of laser-exposed clusters on the basis of an analytical model. Compared to other approaches~\cite{RosPRA97,DitPRA98,LasPRA98,IshPRA00,SiePRL02,BauJPB04,FenPRL07b,PetPP08}, the model is strongly simplified. But as a key benefit, different contributions, such as cluster size distribution, laser intensity distributions in the focus, and ionization saturation can be separated. That allows visualizing their impact on the experimental recoil energy spectra.

\begin{figure}[t]
\centering
	\resizebox{0.6\columnwidth}{!}{\includegraphics{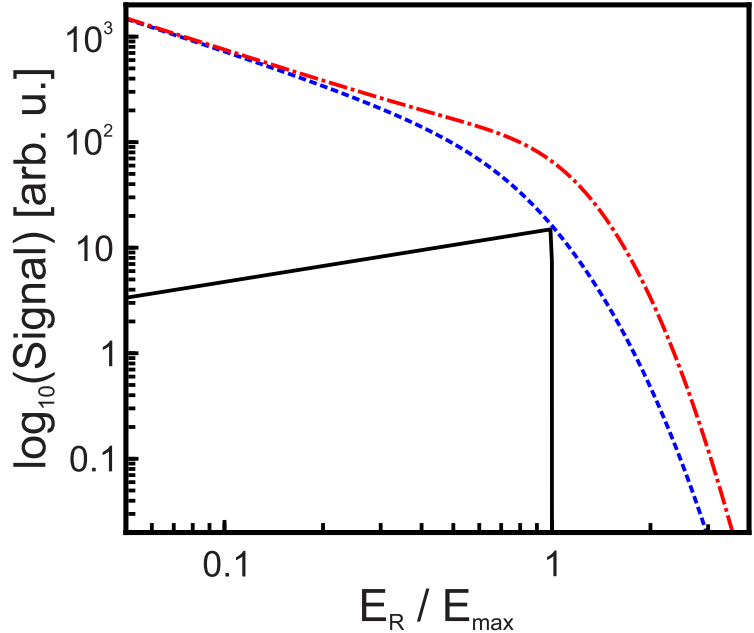}} 
\caption{(\emph{color online}) Ion recoil energy spectra of homogeneously charged Ar$_N$ clusters calculated according to the model of Islam \emph{et al.}~\cite{IslPRA06}. \emph{black solid line:} Single-sized clusters exposed to a given laser intensity. \emph{blue dashed line:} Including a log-normal cluster size distribution and a Gaussian laser beam profile. \emph{red dash-dotted line:} Taking ionization saturation into account. Note, that the energy spectra are presented on double-logarithmic scales.}
\label{fig:KomEPJST20-f1}
\end{figure}

In the model, the cluster is represented as a uniformly charged sphere of constant density and the cluster explosion occurs solely due to Coulomb repulsion forces. The corresponding ion recoil energy spectrum is shown in Fig.~\ref{fig:KomEPJST20-f1} (black solid line). The yield increases with recoil energy $E_R$ up to a maximum energy E$_{\max}$. The low-velocity ions originate from the center of the cluster, whereas ions with $E_R=E_{\max}$ are  released from the surface. In the experiment, however, the spectra differ for several reasons: (i) the size distribution of the clusters in the interaction region and (ii) exposure of clusters to different intensities due to the spatial profile of the focused laser beam. The resulting energy spectrum calculated by considering a log-normal cluster size distribution and a Gaussian laser beam profile is shown in Fig.~\ref{fig:KomEPJST20-f1} (blue dashed line). Large volumes in the focus illuminated by intensities lower than the maximum laser intensity lead to a steep decrease of the yields with energy. The volumetric weighting reflects in a strong contribution of low energy ions to the spectrum. The cluster size distribution results in recoil energies exceeding $E_{\max}$. Finally, consideration of ionization saturation leads to a high energy cut-off (\emph{knee}-feature, see Fig.\,\ref{fig:KomEPJST20-f1} (red dash-dotted line). The simplified modeling can reasonably fit different experimental data~\cite{DitN97,KumPRL01,SakPRA04,KriPRA04,RajRSI11,IwaJPB13}, but the considerable change in the energy spectrum due to the impact of target and laser conditions permits to draw conclusions from the resulting spectrum on the single cluster ion recoil energy distribution.

In the present work, we extend former studies by resolving the impact of the different ion charge states on the recoil energy spectra. The intensities of the laser pulses were chosen in order to monitor the development of the energy spectra with respect to the phenomenon of ionization saturation. The analysis will be conducted in two steps. In order to put the results into the context of previous measurements and the predictions of the model, the individual findings on the recoil energy spectra and the charge-state distribution are presented first. The experimental data show that ionization saturation limits the maximum ion recoil energies already at laser intensities of $10^{15}$\,W/cm$^2$. In the second step, the analysis of the charge-state resolved energy spectra will uncover, that with respect to the highest $q$, the simplified assumption of an expanding homogeneously charged sphere made in the model cannot be retained.

\section{Experimental Setup}
\label{sec:ExpSetup}

\begin{figure}[t]
\centering
	\resizebox{1\columnwidth}{!}{\includegraphics{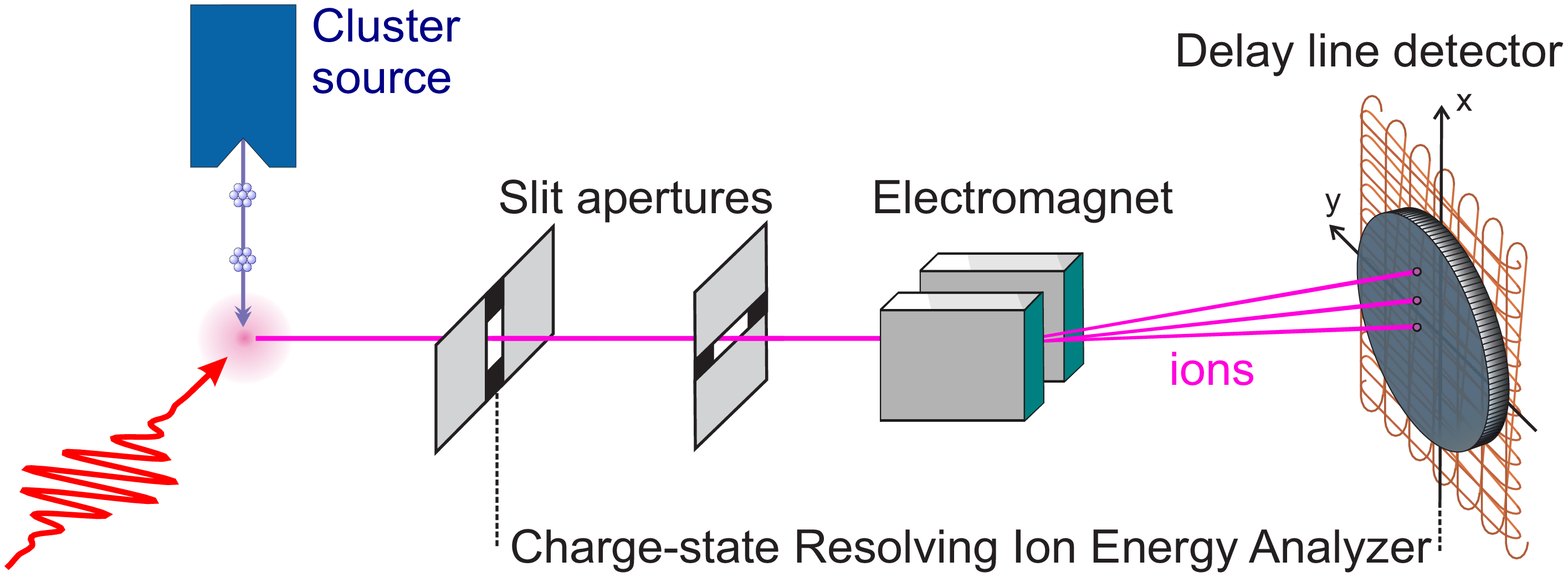}} 
\caption{\emph{(color online)} Experimental setup to measure charge-state selective ion recoil energy spectra from the Coulomb explosion of argon clusters, according to ~\cite{KomRSI16}.}
\label{fig:KomEPJST20-f2}
\end{figure}

The experimental setup to obtain charge-state resolved ion recoil energy spectra from the Coulomb explosion of clusters is schematically shown in Fig.~\ref{fig:KomEPJST20-f2}. Briefly, argon clusters of mean size $\overline{N}$=3800 are produced by supersonic expansion using a pulsed Even-Lavie valve~\cite{PenRSI09}. About 40\,cm behind the nozzle, the particles are exposed to intense 180\,fs near-infrared laser pulses ($\lambda$\,=\,793\,nm). The laser radiation is focused by a lens (f=30\,cm) to a spot diameter of 30\,$\mu$m. An attenuator, based on a $\lambda$/2 plate and a pair of Brewster type polarizers, allows to adjust the intensity between $10^{14}$-$10^{15}$\,W/cm$^2$. In the interaction region, a charge-state resolving ion energy analyzer (CRIEA) is installed~\cite{KomRSI16}. Energetic fragments emitted from the Coulomb explosion are collimated by two slit apertures and pass through a region with a homogeneous magnetic field, and detected by a time- and position-sensitive delay-line detector. The commissioning of the detector system includes a calibration with respect to the actual ion impact position. For this purpose, a specially prepared mask is attached in front of the detector. From the resulting spatially and temporally resolved signals, charge-state resolved ion recoil energy spectra are extracted.

\begin{figure}
\centering
    \resizebox{0.8\columnwidth}{!}{\includegraphics{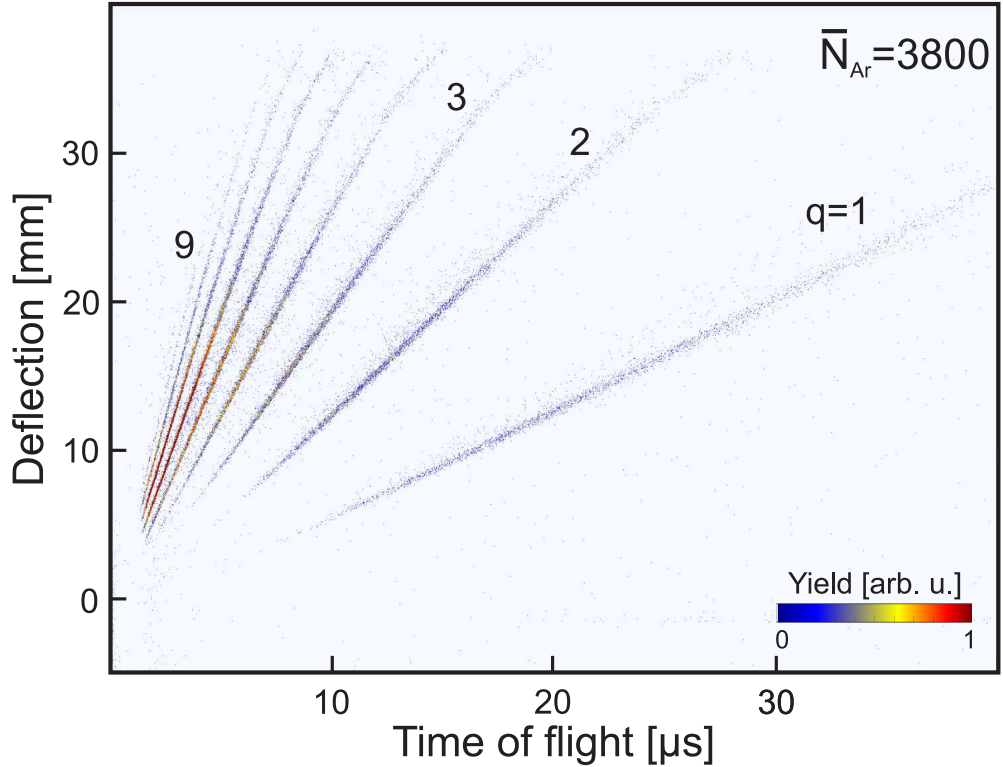}}
\caption{\emph{(color online)} Time-of-flight deflection histogram, obtained from argon clusters with  a mean size of  $\overline{N}$=3800 exposed to laser pulses having peak intensities of 7.3$\cdot$10$^{14}$\,W/cm$^2$. The signal strength is represented by a color scale. Due to different charge states, signals from Ar$^{q+}$ ($q$\,=\,1--9) form lines with different slopes.}
\label{fig:KomEPJST20-f3}
\end{figure}

Compared to other charge-state resolving energy analyzers used in the field like Thomson parabola ~\cite{RajRSI11} or time-of-flight spectrometers with magnetic deflection~\cite{LezPRL98}, CRIEA features (i) a high transmission, since slit apertures are used instead of pinholes. (ii) a high energy resolution even for MeV ions, as instead of the spacial deflection the ion time-of-flight is evaluated. (iii) a delayline detector system, which is characterized by an extended dynamic range with respect to camera based systems and the capability to record data on a shot-to shot basis. (iv) a significant reduction of residual gas signals, since no extraction fields are used. For more detailed information on the  CRIEA design and the procedure to obtain time-of-flight deflection histograms from the delayline detector events, we refer to~\cite{KomRSI16}.

A typical time-of-flight--deflection histogram recorded at a laser intensity of I$_{L}$ = 7.3$\cdot$10$^{14}$\,W/cm$^{2}$ is shown in Fig.~\ref{fig:KomEPJST20-f3}. Charge states from $q$=1 to 9 contribute to the spectrum. The signals from Ar$^{q+}$ for each $q$ form a line due to the different recoil energies $E_R$. Summing up the yields for given $q$ gives the corresponding abundance. Thus, the ion charge-state distribution (CSD) after the Coulomb explosion of the clusters can be obtained. The dependence of the total ion yields as function of time-of-flight $t_{TOF}$ allows to determine the charge integrated ion recoil energy spectrum (IRES), whereas the $t_{TOF}$ of each Ar$^{q+}$ is used to extract the charge-state resolved ion energy spectra (CR-IRES).

\section{Results and discussion}
\label{sec:results}

\begin{figure}[t]
    \centering
	\resizebox{0.7\columnwidth}{!}{\includegraphics{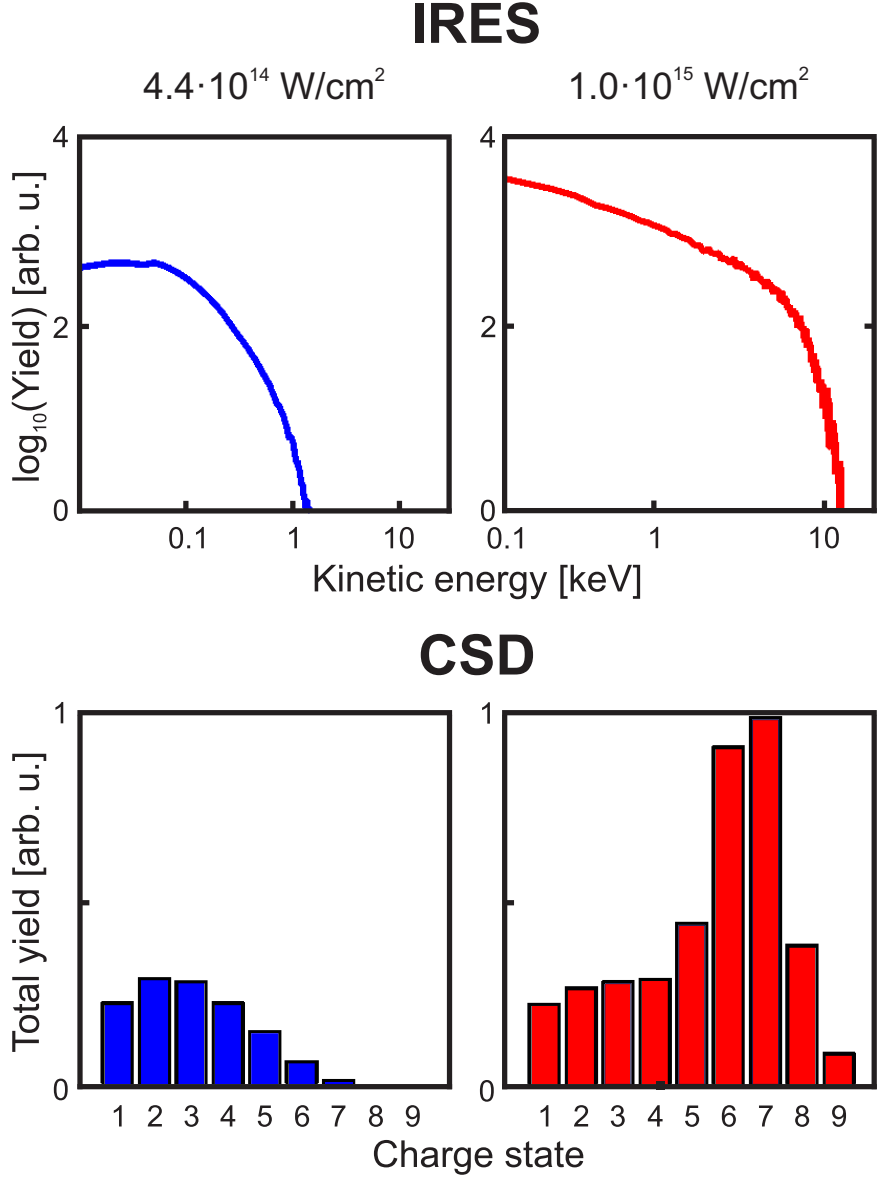}}
	\caption{(\emph{color online}) Ion recoil energy spectra (top panels) and charge-state distributions (bottom panels) at selected laser peak intensities $I_L$= $4.4\cdot10^{14}$\,W/cm$^2$ (left) and  $1.0\cdot10^{15}$\,W/cm$^2$ (right). The IRES are presented on double logarithmic	scales. Note the different energy ranges. The CSD shows an enhanced signal near $q$=7 at $1.0\cdot10^{15}$\,W/cm$^2$  providing evidence for charge-state saturation.}
    \label{fig:KomEPJST20-f4}
\end{figure}

Charge-state integrated recoil ion energy spectra corresponding to selected laser intensities are presented in Fig.~\ref{fig:KomEPJST20-f4} (top panels) on a double logarithmic scale. At $4\cdot10^{14}$\,W/cm$^2$ the IRES spreads from zero to about $E_{\max}$= 2\,keV. Up to about 100\,eV the signal is almost constant and falls off rapidly at higher energies. When increasing the pulse intensity to $10^{15}$\,W/cm$^2$ the recoil energies reach values up to 18\,keV. Hence, doubling the intensity leads to a nearly 10-fold increase in the maximum recoil energy $E_{\max}$.

The substantial increase in $E_R$ between $4\cdot10^{14}$\,W/cm$^2$ and $1\cdot10^{15}$\,W/cm$^2$ can be traced back to resonant charging due to plasmonic excitations~\cite{DitPRA96,KoePRL99}. In the nanoplasma formation, charging of the cluster induces an ion pressure which triggers the Coulomb explosion of the system. The rapid inner and outer ionization~\cite{LasJCP04a} shifts the corresponding Mie plasmon to energies far higher than the laser frequency. In the later expansion, the Mie frequency lowers as a result of the decreasing ion density. Only for the higher laser intensities, the expansion rate is sufficient that the collective mode can effectively adapt to the frequency of the driving laser field. Thus, the efficient charging of the nanoplasma reflects in a substantial increase in $E_R$. Since the ratio between inner and outer ionization reduces with size, we expect that small clusters are mainly responsible for the increase of the recoil energies.  

In addition to the increase of E$_{\max}$ at higher intensities, the envelope of the IRES changes substantially, as shown in Fig.~\ref{fig:KomEPJST20-f4}, top right.  Up to about 8\,keV the yield follows an exponential fall off. But between $E_R$=8 and 20\,keV the yield decreases by several orders of magnitude, giving a pronounced \emph{knee} feature. Additional information on the development of the IRES can be obtained from an analysis of the charge-state distributions, see Fig.~\ref{fig:KomEPJST20-f4} (bottom panels). At $4\cdot10^{14}$\,W/cm$^2$ the CSD extends up to $q$=7 with a maximum at around $q$=2, which suggests that ionisation saturation hardly plays a role. Therefore, the corresponding IRES has to be compared to the result of the model calculation shown in Fig.~\ref{fig:KomEPJST20-f1}, (blue dashed line). Qualitatively, the calculation reproduces the experimental result. However, at low kinetic energies, the IRES, shown in Fig.~\ref{fig:KomEPJST20-f4} (top left), differs from the theoretical result. The difference can be attributed to the reduced detection probability of the multichannel plate detector for ions with a low kinetic energy~\cite{RajPRA12}.     

Ions with charge states up to $q$=9 contribute to the IRES at the higher fluence, see Fig.~\ref{fig:KomEPJST20-f4} (bottom right). But the shape of the CSD changes, too. Up to $q$=4 the Ar$^{q+}$ yields are almost constant, followed by a steep increase, a maximum at Ar$^{7+}$ and  a marked drop beyond. The development of the CSD represents an indication of ionization saturation. For Ar the corresponding feature is expected to stand out through a kink in the CSD at $q$=8 due to the pronounced increase of the ionization potential caused by the $3s^23p^6$ shell closing. The underlying charging mechanism, i.e., laser-assisted electron impact ionization, requires significantly higher energies and thus leads to a strong reduction of the efficiency to produce higher charges states ($q>8$). Hence, for a range of higher laser intensities, no change in the maximum charge state is expected. The anticipated behaviour is in principle reproduced by the measurements. The distribution, however, peaks at $q$ =7, whereas the yield of Ar$^{8+}$ is reduced relative to Ar$^{7+}$. We treat this as an indication of three-body recombination (TBR)~\cite{Bet77} during the nanoplasma disintegration~\cite{FenPRL07b}. Taking ionization saturation into account, the shape of the energy spectrum shown in Fig.~\ref{fig:KomEPJST20-f4} (top right) qualitatively matches the result of the model, see Fig.~\ref{fig:KomEPJST20-f1} (red dash-dotted line).

\begin{figure*}
\centering
    \includegraphics{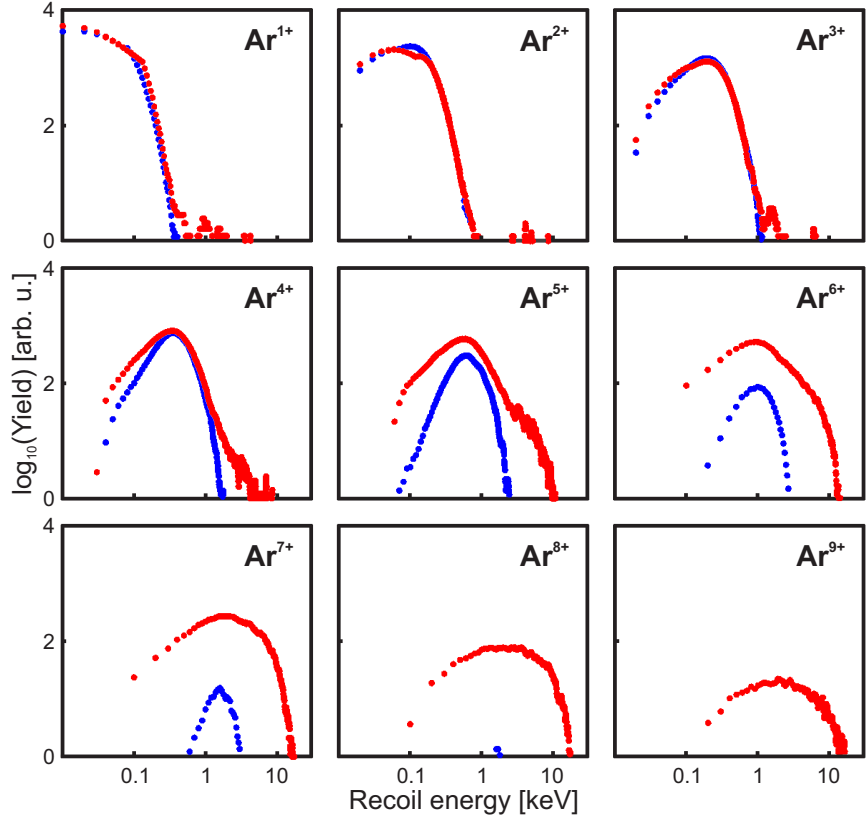}
\caption{\emph{(color online)} Dependence of the recoil energy spectra on the Ar ion charge state recorded at intensities of $4.4\cdot10^{14}$\,W/cm$^2$ (blue) and $1.0\cdot10^{15}$\,W/cm$^2$ (red). In contrast to the energy spectra obtained for  Ar$^{q+}$ ($q$=1-3), the CR-IRES for the higher charge states show pronounced gaps in the ion signals at low kinetic energies.}
\label{fig:KomEPJST20-f5}
\end{figure*}

So far the data have been analyzed by considering either the IRES or the CSD. The ion energy analyzer, however, enables us to extract charge-state resolved ion recoil energy spectra, as shown in Fig.~\ref{fig:KomEPJST20-f5} for both laser intensities. In general, none of the spectra is similar to the IRES, as presented in Fig.~\ref{fig:KomEPJST20-f4}. For Ar$^{1+}$, the recoil energies range from zero up to about 1\,keV. With increasing $q$, the spectra develop into distinct peaks with the maxima shifting towards higher E$_R$. Notably, no ions with low recoil energies are detected for higher charge states. In addition, the energy spectra overlap for given $q$, although the laser intensities differ by more than a factor of two. Most likely, this effect arises as a result of focal averaging~\cite{DoeEPJD07}. When increasing the laser power an additional volume is illuminated by the higher $I_L$, while the volume illuminated by lower intensities remains constant. Hence, if ions of selected Ar$^{q+}$ are not emitted from regions of higher $I_L$, no change in the yields is expected. The corresponding behavior is observed for Ar$^{1+}$ to Ar$^{3+}$, see Fig.~\ref{fig:KomEPJST20-f5}, which suggests that these charge states are produced mainly in the low-intensity regions of the focus. For $q$=4-7, focal averaging may be responsible for the peaks obtained at around 1 keV, as the position of the maxima are similar at both intensities.

Whereas the CR-IRES of Ar$^{1+}$-Ar$^{3+}$ show strong contributions from slow-moving ions, the spectra beyond exhibit pronounced low energy cut-offs. For example, the spectrum of Ar$^{7+}$ taken at $I_L=4.4\cdot10^{14}$\,W/cm$^2$ shows recoil energies up to $E_{\max}$= 3.1\,keV. At the same time, the energy distribution exhibits a sudden onset of the signal at $E_{\min}$=600\,eV. This behavior and the development of the distinct peaks with increasing $q$, cannot obviously be represented as being the sum of $q$-dependent recoil energy spectra stemming from homogeneously exploding charged spheres as assumed in~\cite{IslPRA06}. Since ions at the surface of the cluster gain the highest kinetic energy, the experimental observation implies, that the higher charged ions reside predominantly near the cluster surface. Since cluster ionization is mainly caused by laser-assisted electron impact ionization, an excessive charging of surface atoms compared to the interior is not expected. Hence, in order to reproduce the experimental results, non-trivial spatial charge distributions have to be assumed. Inhomogeneities of the spatial charge distribution as a result of the oscillating quasifree electron cloud may be responsible for the effect~\cite{KumPRA02,KumPRL01,SprPRA00}. But an impact of the electron motion has only been observed for the highest recoil energies~\cite{KumPRL01}. Plasmon-induced charge fluctuations can, therefore, be ruled out to explain the small contribution of low energy ions in the spectra of Ar$^{q+}$. 

Most probably, TBR is responsible and plays an essential role in the temporal development of the charge-state distribution~\cite{FenPRL07b,ThoPRL12}. Initially, laser-assisted avalanche charging of the cluster produces ions in high charge states. However, in the expansion of the nanoplasma, ions primary near the cluster core recombine with quasi-free electrons, whereas for Ar$^{q+}$ near the plasma surface TBR is less effective. This results in a spatial imbalance of the charge-state distribution, i.e.,  surface ions will have a higher charge state and experience a larger Coulomb pressure.

Note, that the CSD spectrum comprises information about the contribution of electron-ion recombination to the final ion charge state. Hence, the full expansion period is mapped without temporal resolution. In contrast, the CR-IRES spectra are sensitive to the period when the ions accumulate most of their kinetic energy, which roughly ties to the laser pulse duration. The low-energy cutoffs indicate that TBR contributes markedly from the very beginning of the interaction. Although the nanoplasma temperatures during the laser pulse impact are expected to be high, hence suppressing TBR, the CR-IRES suggest, that recombination has to be taken into account on all timescales. A related computational study on the recoil energy distributions is appealing to obtain more detailed information. Conducting corresponding simulations, however, are beyond the scope of the contribution.

The charge-state distribution of larger argon clusters, i.e. Ar$_N$, $N=36\,000$ at intensities of $7\cdot10^{15}$\,W/cm$^2$ has been studied by Rajeev et al.~\cite{RajPRA12}. The authors obtained an envelope of the CSD similar to the one observed in the present work at $I_L\,=10^{15}$\,W/cm$^2$, but the CSD shift and peaks at $q$=8 due to the larger size and the higher pulse intensity. Hence, the observed trend is found to be in accordance with our measurements. However, in contrast to our study, those experiments show no signals from the lower charge states, i.e., $q\leq$4. With respect to the slight increase in the intensity conditions, the absence of ions in low $q$-states in the CSD remains surprising. Irrespective on the chosen laser intensity, one would expect to obtain ion signals from the lower charge states due to focal averaging. Hence, in order to resolve the contradictory results e.g., intensity-selective scanning experiments have to be conducted~\cite{DoePRL10}. In view of the need to finally compare the results to molecular dynamics calculations, such a treatment is appealing, since the simulations, are typically conducted at only a single laser intensity~\cite{LasPRA00}.      

Finally, the resulting spectra have to be linked to simulations on Ar$_{40\,000}$~\cite{KriJPB06}. At $I_L=10^{15}$\,W/cm$^2$, the calculations show that the CSD peaks at $q$=6, whereas at $3\cdot10^{15}$\,W/cm$^2$ the maximum of the distribution shifts to $q$=7. The computational result concurs with our findings. However, the corresponding energy spectra show that the \emph{knee} energy exceeds the experimental value obtained in the present work by more than a factor of four. As a possible cause, a lower plasma  electron temperature $T_e$ could be responsible as TBR strongly depends on $T_e$, i.e., ($T_e^{-9/2}$)~\cite{ManPRev69}. Further, the size of the argon particles may play a role. According to the model of Islam, the lower \emph{knee}-energy can be explained by the smaller clusters exposed to the laser field. In addition, one can expect that as function of cluster radius, the average nanoplasma charge state decreases. Hence, intensity dependent measurements in this regime are then again appealing, since, e.g., the CSD has been found to be quite sensitive to small changes in laser power. Moreover, the CR-IRES results will allow for a more in-depth analysis. 

\section*{Conclusions}
Recoil ion energy distributions from the Coulomb explosion of small Ar clusters exposed to strong laser pulses have been studied by charge and energy-resolved spectroscopy. The recoil energy spectra taken at the highest laser intensities of $10^{15}$\,W/\,cm$^2$ exhibit a high-energy cut-off. The corresponding charge-state distribution indicates, that the feature stems from ionization saturation. The envelope of the recoil energy spectrum can be well-described by a model, which takes into account ionization saturation as well as details of cluster size and laser intensity distributions. At lower laser intensities, the in-depth information provided by resolving the contributions of the individual charge states to the ion energy spectra reveal distinct low-energy gaps for the higher charge states. Since the ions gain a substantial fraction of their recoil energy within a short period of time, the charge-state resolved energy spectra are sensitive to the nanoplasma condition in the early expansion period.  The experiments give evidence that the corresponding charge-state distribution developing during the laser pulse impact has to be assumed to be inhomogeneous in order to match the experimental results. The presence of low-energy gaps point out the relevance of three-body recombination already during the laser pulse impact.

\section*{Acknowledgments}
The Deutsche Forschungsgemeinschaft (TI210-8) is gratefully acknowledged for financial support.

\section*{Author contributions}
D.K. and J.T. conceived the experiment; D.K. and L.K. performed the measurements; D.K. analyzed the data; all authors contributed to the analysis; L.K., D.K., K.H.M.B., and J.T. wrote the manuscript.


\begin{thebibliography}{50}%
\makeatletter
\providecommand \@ifxundefined [1]{%
 \@ifx{#1\undefined}
}%
\providecommand \@ifnum [1]{%
 \ifnum #1\expandafter \@firstoftwo
 \else \expandafter \@secondoftwo
 \fi
}%
\providecommand \@ifx [1]{%
 \ifx #1\expandafter \@firstoftwo
 \else \expandafter \@secondoftwo
 \fi
}%
\providecommand \natexlab [1]{#1}%
\providecommand \enquote  [1]{``#1''}%
\providecommand \bibnamefont  [1]{#1}%
\providecommand \bibfnamefont [1]{#1}%
\providecommand \citenamefont [1]{#1}%
\providecommand \href@noop [0]{\@secondoftwo}%
\providecommand \href [0]{\begingroup \@sanitize@url \@href}%
\providecommand \@href[1]{\@@startlink{#1}\@@href}%
\providecommand \@@href[1]{\endgroup#1\@@endlink}%
\providecommand \@sanitize@url [0]{\catcode `\\12\catcode `\$12\catcode
  `\&12\catcode `\#12\catcode `\^12\catcode `\_12\catcode `\%12\relax}%
\providecommand \@@startlink[1]{}%
\providecommand \@@endlink[0]{}%
\providecommand \url  [0]{\begingroup\@sanitize@url \@url }%
\providecommand \@url [1]{\endgroup\@href {#1}{\urlprefix }}%
\providecommand \urlprefix  [0]{URL }%
\providecommand \Eprint [0]{\href }%
\providecommand \doibase [0]{http://dx.doi.org/}%
\providecommand \selectlanguage [0]{\@gobble}%
\providecommand \bibinfo  [0]{\@secondoftwo}%
\providecommand \bibfield  [0]{\@secondoftwo}%
\providecommand \translation [1]{[#1]}%
\providecommand \BibitemOpen [0]{}%
\providecommand \bibitemStop [0]{}%
\providecommand \bibitemNoStop [0]{.\EOS\space}%
\providecommand \EOS [0]{\spacefactor3000\relax}%
\providecommand \BibitemShut  [1]{\csname bibitem#1\endcsname}%
\let\auto@bib@innerbib\@empty
\bibitem [{\citenamefont {Esarey}\ \emph {et~al.}(1996)\citenamefont {Esarey},
  \citenamefont {Sprangle}, \citenamefont {Krall},\ and\ \citenamefont
  {Ting}}]{EsaTPS96}%
  \BibitemOpen
  \bibfield  {author} {\bibinfo {author} {\bibfnamefont {E.}~\bibnamefont
  {Esarey}}, \bibinfo {author} {\bibfnamefont {P.}~\bibnamefont {Sprangle}},
  \bibinfo {author} {\bibfnamefont {J.}~\bibnamefont {Krall}}, \ and\ \bibinfo
  {author} {\bibfnamefont {A.}~\bibnamefont {Ting}},\ }\href@noop {} {\bibfield
   {journal} {\bibinfo  {journal} {Trans. Plas. Sci.}\ }\textbf {\bibinfo
  {volume} {24}},\ \bibinfo {pages} {252} (\bibinfo {year} {1996})}\BibitemShut
  {NoStop}%
\bibitem [{\citenamefont {Kmetec}\ \emph {et~al.}(1992)\citenamefont {Kmetec},
  \citenamefont {Macklin}, \citenamefont {Lemoff}, \citenamefont {Brown},\ and\
  \citenamefont {Harris}}]{KmePRL92}%
  \BibitemOpen
  \bibfield  {author} {\bibinfo {author} {\bibfnamefont {J.}~\bibnamefont
  {Kmetec}}, \bibinfo {author} {\bibfnamefont {J.}~\bibnamefont {Macklin}},
  \bibinfo {author} {\bibfnamefont {B.}~\bibnamefont {Lemoff}}, \bibinfo
  {author} {\bibfnamefont {G.}~\bibnamefont {Brown}}, \ and\ \bibinfo {author}
  {\bibfnamefont {S.}~\bibnamefont {Harris}},\ }\href@noop {} {\bibfield
  {journal} {\bibinfo  {journal} {Phys. Rev. Lett.}\ }\textbf {\bibinfo
  {volume} {68}},\ \bibinfo {pages} {1527} (\bibinfo {year}
  {1992})}\BibitemShut {NoStop}%
\bibitem [{\citenamefont {Krainov}\ and\ \citenamefont
  {Smirnov}(2002)}]{KraPR02}%
  \BibitemOpen
  \bibfield  {author} {\bibinfo {author} {\bibfnamefont {V.}~\bibnamefont
  {Krainov}}\ and\ \bibinfo {author} {\bibfnamefont {M.}~\bibnamefont
  {Smirnov}},\ }\href@noop {} {\bibfield  {journal} {\bibinfo  {journal} {Phys.
  Rep.}\ }\textbf {\bibinfo {volume} {370}},\ \bibinfo {pages} {237} (\bibinfo
  {year} {2002})}\BibitemShut {NoStop}%
\bibitem [{\citenamefont {Saalmann}\ \emph {et~al.}(2006)\citenamefont
  {Saalmann}, \citenamefont {Siedschlag},\ and\ \citenamefont
  {Rost}}]{SaaJPB06}%
  \BibitemOpen
  \bibfield  {author} {\bibinfo {author} {\bibfnamefont {U.}~\bibnamefont
  {Saalmann}}, \bibinfo {author} {\bibfnamefont {C.}~\bibnamefont
  {Siedschlag}}, \ and\ \bibinfo {author} {\bibfnamefont {J.~M.}\ \bibnamefont
  {Rost}},\ }\href@noop {} {\bibfield  {journal} {\bibinfo  {journal} {J. Phys.
  B}\ }\textbf {\bibinfo {volume} {39}},\ \bibinfo {pages} {R39} (\bibinfo
  {year} {2006})}\BibitemShut {NoStop}%
\bibitem [{\citenamefont {Fennel}\ \emph {et~al.}(2010)\citenamefont {Fennel},
  \citenamefont {Meiwes-Broer}, \citenamefont {Tiggesb\"aumker}, \citenamefont
  {Reinhard}, \citenamefont {Dinh},\ and\ \citenamefont {Suraud}}]{FenRMP10}%
  \BibitemOpen
  \bibfield  {author} {\bibinfo {author} {\bibfnamefont {T.}~\bibnamefont
  {Fennel}}, \bibinfo {author} {\bibfnamefont {K.-H.}\ \bibnamefont
  {Meiwes-Broer}}, \bibinfo {author} {\bibfnamefont {J.}~\bibnamefont
  {Tiggesb\"aumker}}, \bibinfo {author} {\bibfnamefont {P.-G.}\ \bibnamefont
  {Reinhard}}, \bibinfo {author} {\bibfnamefont {P.~M.}\ \bibnamefont {Dinh}},
  \ and\ \bibinfo {author} {\bibfnamefont {E.}~\bibnamefont {Suraud}},\
  }\href@noop {} {\bibfield  {journal} {\bibinfo  {journal} {Rev. Mod. Phys.}\
  }\textbf {\bibinfo {volume} {82}},\ \bibinfo {pages} {1793} (\bibinfo {year}
  {2010})}\BibitemShut {NoStop}%
\bibitem [{\citenamefont {Ditmire}\ \emph {et~al.}(1996)\citenamefont
  {Ditmire}, \citenamefont {Donnelly}, \citenamefont {Rubenchik}, \citenamefont
  {Falcone},\ and\ \citenamefont {Perry}}]{DitPRA96}%
  \BibitemOpen
  \bibfield  {author} {\bibinfo {author} {\bibfnamefont {T.}~\bibnamefont
  {Ditmire}}, \bibinfo {author} {\bibfnamefont {T.}~\bibnamefont {Donnelly}},
  \bibinfo {author} {\bibfnamefont {A.~M.}\ \bibnamefont {Rubenchik}}, \bibinfo
  {author} {\bibfnamefont {R.~W.}\ \bibnamefont {Falcone}}, \ and\ \bibinfo
  {author} {\bibfnamefont {M.~D.}\ \bibnamefont {Perry}},\ }\href@noop {}
  {\bibfield  {journal} {\bibinfo  {journal} {Phys. Rev. A}\ }\textbf {\bibinfo
  {volume} {53}},\ \bibinfo {pages} {3379} (\bibinfo {year}
  {1996})}\BibitemShut {NoStop}%
\bibitem [{\citenamefont {Saalmann}\ and\ \citenamefont
  {Rost}(2003)}]{SaaPRL03}%
  \BibitemOpen
  \bibfield  {author} {\bibinfo {author} {\bibfnamefont {U.}~\bibnamefont
  {Saalmann}}\ and\ \bibinfo {author} {\bibfnamefont {J.-M.}\ \bibnamefont
  {Rost}},\ }\href@noop {} {\bibfield  {journal} {\bibinfo  {journal} {Phys.
  Rev. Lett.}\ }\textbf {\bibinfo {volume} {91}},\ \bibinfo {pages} {223401}
  (\bibinfo {year} {2003})}\BibitemShut {NoStop}%
\bibitem [{\citenamefont {Fennel}\ \emph
  {et~al.}(2007{\natexlab{a}})\citenamefont {Fennel}, \citenamefont
  {D\"oppner}, \citenamefont {Passig}, \citenamefont {Schaal}, \citenamefont
  {Tiggesb\"aumker},\ and\ \citenamefont {Meiwes-Broer}}]{FenPRL07a}%
  \BibitemOpen
  \bibfield  {author} {\bibinfo {author} {\bibfnamefont {T.}~\bibnamefont
  {Fennel}}, \bibinfo {author} {\bibfnamefont {T.}~\bibnamefont {D\"oppner}},
  \bibinfo {author} {\bibfnamefont {J.}~\bibnamefont {Passig}}, \bibinfo
  {author} {\bibfnamefont {C.}~\bibnamefont {Schaal}}, \bibinfo {author}
  {\bibfnamefont {J.}~\bibnamefont {Tiggesb\"aumker}}, \ and\ \bibinfo {author}
  {\bibfnamefont {K.-H.}\ \bibnamefont {Meiwes-Broer}},\ }\href@noop {}
  {\bibfield  {journal} {\bibinfo  {journal} {Phys. Rev. Lett.}\ }\textbf
  {\bibinfo {volume} {98}},\ \bibinfo {pages} {143401} (\bibinfo {year}
  {2007}{\natexlab{a}})}\BibitemShut {NoStop}%
\bibitem [{\citenamefont {Snyder}\ \emph {et~al.}(1996)\citenamefont {Snyder},
  \citenamefont {Buzza},\ and\ \citenamefont {{Castleman Jr.}}}]{SnyPRL96}%
  \BibitemOpen
  \bibfield  {author} {\bibinfo {author} {\bibfnamefont {E.~M.}\ \bibnamefont
  {Snyder}}, \bibinfo {author} {\bibfnamefont {S.~A.}\ \bibnamefont {Buzza}}, \
  and\ \bibinfo {author} {\bibfnamefont {A.~W.}\ \bibnamefont {{Castleman
  Jr.}}},\ }\href@noop {} {\bibfield  {journal} {\bibinfo  {journal} {Phys. Rev
  Lett.}\ }\textbf {\bibinfo {volume} {77}},\ \bibinfo {pages} {3347} (\bibinfo
  {year} {1996})}\BibitemShut {NoStop}%
\bibitem [{\citenamefont {Lezius}\ \emph {et~al.}(1998)\citenamefont {Lezius},
  \citenamefont {Dobosz}, \citenamefont {Normand},\ and\ \citenamefont
  {Schmidt}}]{LezPRL98}%
  \BibitemOpen
  \bibfield  {author} {\bibinfo {author} {\bibfnamefont {M.}~\bibnamefont
  {Lezius}}, \bibinfo {author} {\bibfnamefont {S.}~\bibnamefont {Dobosz}},
  \bibinfo {author} {\bibfnamefont {D.}~\bibnamefont {Normand}}, \ and\
  \bibinfo {author} {\bibfnamefont {M.}~\bibnamefont {Schmidt}},\ }\href@noop
  {} {\bibfield  {journal} {\bibinfo  {journal} {Phys. Rev. Lett.}\ }\textbf
  {\bibinfo {volume} {80}},\ \bibinfo {pages} {261} (\bibinfo {year}
  {1998})}\BibitemShut {NoStop}%
\bibitem [{\citenamefont {Springate}\ \emph {et~al.}(2003)\citenamefont
  {Springate}, \citenamefont {Aseyev}, \citenamefont {Zamith},\ and\
  \citenamefont {Vrakking}}]{SprPRA03}%
  \BibitemOpen
  \bibfield  {author} {\bibinfo {author} {\bibfnamefont {E.}~\bibnamefont
  {Springate}}, \bibinfo {author} {\bibfnamefont {S.~A.}\ \bibnamefont
  {Aseyev}}, \bibinfo {author} {\bibfnamefont {S.}~\bibnamefont {Zamith}}, \
  and\ \bibinfo {author} {\bibfnamefont {M.~J.~J.}\ \bibnamefont {Vrakking}},\
  }\href@noop {} {\bibfield  {journal} {\bibinfo  {journal} {Phys. Rev. A}\
  }\textbf {\bibinfo {volume} {68}},\ \bibinfo {pages} {053201} (\bibinfo
  {year} {2003})}\BibitemShut {NoStop}%
\bibitem [{\citenamefont {Passig}\ \emph {et~al.}(2017)\citenamefont {Passig},
  \citenamefont {Zherebtsov}, \citenamefont {Irsig}, \citenamefont {Arbeiter},
  \citenamefont {Peltz}, \citenamefont {G\"ode}, \citenamefont {Skruszewicz},
  \citenamefont {Meiwes-Broer}, \citenamefont {Tiggesb\"aumker}, \citenamefont
  {Kling},\ and\ \citenamefont {Fennel}}]{PasNCom17}%
  \BibitemOpen
  \bibfield  {author} {\bibinfo {author} {\bibfnamefont {J.}~\bibnamefont
  {Passig}}, \bibinfo {author} {\bibfnamefont {S.}~\bibnamefont {Zherebtsov}},
  \bibinfo {author} {\bibfnamefont {R.}~\bibnamefont {Irsig}}, \bibinfo
  {author} {\bibfnamefont {M.}~\bibnamefont {Arbeiter}}, \bibinfo {author}
  {\bibfnamefont {C.}~\bibnamefont {Peltz}}, \bibinfo {author} {\bibfnamefont
  {S.}~\bibnamefont {G\"ode}}, \bibinfo {author} {\bibfnamefont
  {S.}~\bibnamefont {Skruszewicz}}, \bibinfo {author} {\bibfnamefont {K.-H.}\
  \bibnamefont {Meiwes-Broer}}, \bibinfo {author} {\bibfnamefont
  {J.}~\bibnamefont {Tiggesb\"aumker}}, \bibinfo {author} {\bibfnamefont
  {M.}~\bibnamefont {Kling}}, \ and\ \bibinfo {author} {\bibfnamefont
  {T.}~\bibnamefont {Fennel}},\ }\href {\doibase 10.1038/s41467-017-01286-w}
  {\bibfield  {journal} {\bibinfo  {journal} {Nature Commun.}\ }\textbf
  {\bibinfo {volume} {8}},\ \bibinfo {pages} {1181} (\bibinfo {year}
  {2017})}\BibitemShut {NoStop}%
\bibitem [{\citenamefont {Hirokane}\ \emph {et~al.}(2004)\citenamefont
  {Hirokane}, \citenamefont {Shimizu}, \citenamefont {Hashida}, \citenamefont
  {Okada}, \citenamefont {Okihara}, \citenamefont {Sato}, \citenamefont
  {Iida},\ and\ \citenamefont {Sakabe}}]{HirPRA04}%
  \BibitemOpen
  \bibfield  {author} {\bibinfo {author} {\bibfnamefont {M.}~\bibnamefont
  {Hirokane}}, \bibinfo {author} {\bibfnamefont {S.}~\bibnamefont {Shimizu}},
  \bibinfo {author} {\bibfnamefont {M.}~\bibnamefont {Hashida}}, \bibinfo
  {author} {\bibfnamefont {S.}~\bibnamefont {Okada}}, \bibinfo {author}
  {\bibfnamefont {S.}~\bibnamefont {Okihara}}, \bibinfo {author} {\bibfnamefont
  {F.}~\bibnamefont {Sato}}, \bibinfo {author} {\bibfnamefont {T.}~\bibnamefont
  {Iida}}, \ and\ \bibinfo {author} {\bibfnamefont {S.}~\bibnamefont
  {Sakabe}},\ }\href@noop {} {\bibfield  {journal} {\bibinfo  {journal} {Phys.
  Rev. Lett.}\ }\textbf {\bibinfo {volume} {69}},\ \bibinfo {pages} {063201}
  (\bibinfo {year} {2004})}\BibitemShut {NoStop}%
\bibitem [{\citenamefont {Rajeev}\ \emph {et~al.}(2013)\citenamefont {Rajeev},
  \citenamefont {Trivikram}, \citenamefont {Rishad}, \citenamefont {Narayanan},
  \citenamefont {Krishnakumar},\ and\ \citenamefont
  {Krishnamurthy}}]{RajNPhys13}%
  \BibitemOpen
  \bibfield  {author} {\bibinfo {author} {\bibfnamefont {R.}~\bibnamefont
  {Rajeev}}, \bibinfo {author} {\bibfnamefont {T.~M.}\ \bibnamefont
  {Trivikram}}, \bibinfo {author} {\bibfnamefont {K.}~\bibnamefont {Rishad}},
  \bibinfo {author} {\bibfnamefont {V.}~\bibnamefont {Narayanan}}, \bibinfo
  {author} {\bibfnamefont {E.}~\bibnamefont {Krishnakumar}}, \ and\ \bibinfo
  {author} {\bibfnamefont {M.}~\bibnamefont {Krishnamurthy}},\ }\href@noop {}
  {\bibfield  {journal} {\bibinfo  {journal} {Nature Phys.}\ }\textbf {\bibinfo
  {volume} {9}},\ \bibinfo {pages} {185} (\bibinfo {year} {2013})}\BibitemShut
  {NoStop}%
\bibitem [{\citenamefont {Rupp}\ \emph {et~al.}(2016)\citenamefont {Rupp},
  \citenamefont {Fl\"uckiger}, \citenamefont {Adolph}, \citenamefont
  {Gorkhover}, \citenamefont {Krikunova}, \citenamefont {M\"uller},
  \citenamefont {M\"uller}, \citenamefont {Oelze}, \citenamefont {Ovcharenko},
  \citenamefont {R\"oben}, \citenamefont {Sauppe}, \citenamefont {Schorb},
  \citenamefont {Wolter}, \citenamefont {Mitzner}, \citenamefont {W\"ostmann},
  \citenamefont {Roling}, \citenamefont {Harmand}, \citenamefont {Treusch},
  \citenamefont {Arbeiter}, \citenamefont {Fennel}, \citenamefont {Bostedt},\
  and\ \citenamefont {M\"oller}}]{RupPRL16}%
  \BibitemOpen
  \bibfield  {author} {\bibinfo {author} {\bibfnamefont {D.}~\bibnamefont
  {Rupp}}, \bibinfo {author} {\bibfnamefont {L.}~\bibnamefont {Fl\"uckiger}},
  \bibinfo {author} {\bibfnamefont {M.}~\bibnamefont {Adolph}}, \bibinfo
  {author} {\bibfnamefont {T.}~\bibnamefont {Gorkhover}}, \bibinfo {author}
  {\bibfnamefont {M.}~\bibnamefont {Krikunova}}, \bibinfo {author}
  {\bibfnamefont {J.~P.}\ \bibnamefont {M\"uller}}, \bibinfo {author}
  {\bibfnamefont {M.}~\bibnamefont {M\"uller}}, \bibinfo {author}
  {\bibfnamefont {T.}~\bibnamefont {Oelze}}, \bibinfo {author} {\bibfnamefont
  {Y.}~\bibnamefont {Ovcharenko}}, \bibinfo {author} {\bibfnamefont
  {B.}~\bibnamefont {R\"oben}}, \bibinfo {author} {\bibfnamefont
  {M.}~\bibnamefont {Sauppe}}, \bibinfo {author} {\bibfnamefont
  {S.}~\bibnamefont {Schorb}}, \bibinfo {author} {\bibfnamefont
  {D.}~\bibnamefont {Wolter}}, \bibinfo {author} {\bibfnamefont
  {R.}~\bibnamefont {Mitzner}}, \bibinfo {author} {\bibfnamefont
  {M.}~\bibnamefont {W\"ostmann}}, \bibinfo {author} {\bibfnamefont
  {S.}~\bibnamefont {Roling}}, \bibinfo {author} {\bibfnamefont
  {M.}~\bibnamefont {Harmand}}, \bibinfo {author} {\bibfnamefont
  {R.}~\bibnamefont {Treusch}}, \bibinfo {author} {\bibfnamefont
  {M.}~\bibnamefont {Arbeiter}}, \bibinfo {author} {\bibfnamefont
  {T.}~\bibnamefont {Fennel}}, \bibinfo {author} {\bibfnamefont
  {C.}~\bibnamefont {Bostedt}}, \ and\ \bibinfo {author} {\bibfnamefont
  {T.}~\bibnamefont {M\"oller}},\ }\href {\doibase
  10.1103/PhysRevLett.117.153401} {\bibfield  {journal} {\bibinfo  {journal}
  {Phys. Rev. Lett.}\ }\textbf {\bibinfo {volume} {117}},\ \bibinfo {pages}
  {153401} (\bibinfo {year} {2016})}\BibitemShut {NoStop}%
\bibitem [{\citenamefont {Kumagai}\ \emph {et~al.}(2018)\citenamefont
  {Kumagai}, \citenamefont {Fukuzawa}, \citenamefont {Motomura}, \citenamefont
  {Iablonskyi}, \citenamefont {Nagaya}, \citenamefont {Wada}, \citenamefont
  {Ito}, \citenamefont {Takanashi}, \citenamefont {Sakakibara}, \citenamefont
  {You}, \citenamefont {Nishiyama}, \citenamefont {Asa}, \citenamefont {Sato},
  \citenamefont {Umemoto}, \citenamefont {Kariyazono}, \citenamefont {Kukk},
  \citenamefont {Kooser}, \citenamefont {Nicolas}, \citenamefont {Miron},
  \citenamefont {Asavei}, \citenamefont {Neagu}, \citenamefont {Sch\"offler},
  \citenamefont {Kastirke}, \citenamefont {Liu}, \citenamefont {Owada},
  \citenamefont {Katayama}, \citenamefont {Togashi}, \citenamefont {Tono},
  \citenamefont {Yabashi}, \citenamefont {Golubev}, \citenamefont {Gokhberg},
  \citenamefont {Cederbaum}, \citenamefont {Kuleff},\ and\ \citenamefont
  {Ueda}}]{KumPRX18}%
  \BibitemOpen
  \bibfield  {author} {\bibinfo {author} {\bibfnamefont {Y.}~\bibnamefont
  {Kumagai}}, \bibinfo {author} {\bibfnamefont {H.}~\bibnamefont {Fukuzawa}},
  \bibinfo {author} {\bibfnamefont {K.}~\bibnamefont {Motomura}}, \bibinfo
  {author} {\bibfnamefont {D.}~\bibnamefont {Iablonskyi}}, \bibinfo {author}
  {\bibfnamefont {K.}~\bibnamefont {Nagaya}}, \bibinfo {author} {\bibfnamefont
  {S.~I.}\ \bibnamefont {Wada}}, \bibinfo {author} {\bibfnamefont
  {Y.}~\bibnamefont {Ito}}, \bibinfo {author} {\bibfnamefont {T.}~\bibnamefont
  {Takanashi}}, \bibinfo {author} {\bibfnamefont {Y.}~\bibnamefont
  {Sakakibara}}, \bibinfo {author} {\bibfnamefont {D.}~\bibnamefont {You}},
  \bibinfo {author} {\bibfnamefont {T.}~\bibnamefont {Nishiyama}}, \bibinfo
  {author} {\bibfnamefont {K.}~\bibnamefont {Asa}}, \bibinfo {author}
  {\bibfnamefont {Y.}~\bibnamefont {Sato}}, \bibinfo {author} {\bibfnamefont
  {T.}~\bibnamefont {Umemoto}}, \bibinfo {author} {\bibfnamefont
  {K.}~\bibnamefont {Kariyazono}}, \bibinfo {author} {\bibfnamefont
  {E.}~\bibnamefont {Kukk}}, \bibinfo {author} {\bibfnamefont {K.}~\bibnamefont
  {Kooser}}, \bibinfo {author} {\bibfnamefont {C.}~\bibnamefont {Nicolas}},
  \bibinfo {author} {\bibfnamefont {C.}~\bibnamefont {Miron}}, \bibinfo
  {author} {\bibfnamefont {T.}~\bibnamefont {Asavei}}, \bibinfo {author}
  {\bibfnamefont {L.}~\bibnamefont {Neagu}}, \bibinfo {author} {\bibfnamefont
  {M.~S.}\ \bibnamefont {Sch\"offler}}, \bibinfo {author} {\bibfnamefont
  {G.}~\bibnamefont {Kastirke}}, \bibinfo {author} {\bibfnamefont {X.-J.}\
  \bibnamefont {Liu}}, \bibinfo {author} {\bibfnamefont {S.}~\bibnamefont
  {Owada}}, \bibinfo {author} {\bibfnamefont {T.}~\bibnamefont {Katayama}},
  \bibinfo {author} {\bibfnamefont {T.}~\bibnamefont {Togashi}}, \bibinfo
  {author} {\bibfnamefont {K.}~\bibnamefont {Tono}}, \bibinfo {author}
  {\bibfnamefont {M.}~\bibnamefont {Yabashi}}, \bibinfo {author} {\bibfnamefont
  {N.~V.}\ \bibnamefont {Golubev}}, \bibinfo {author} {\bibfnamefont
  {K.}~\bibnamefont {Gokhberg}}, \bibinfo {author} {\bibfnamefont {L.~S.}\
  \bibnamefont {Cederbaum}}, \bibinfo {author} {\bibfnamefont {A.~I.}\
  \bibnamefont {Kuleff}}, \ and\ \bibinfo {author} {\bibfnamefont
  {K.}~\bibnamefont {Ueda}},\ }\href {\doibase 10.1103/PhysRevX.8.031034}
  {\bibfield  {journal} {\bibinfo  {journal} {Phys. Rev. X}\ }\textbf {\bibinfo
  {volume} {8}},\ \bibinfo {pages} {031034} (\bibinfo {year}
  {2018})}\BibitemShut {NoStop}%
\bibitem [{\citenamefont {Zhang}\ \emph {et~al.}(2020)\citenamefont {Zhang},
  \citenamefont {Yao},\ and\ \citenamefont {Kong}}]{ZhaJPCL20}%
  \BibitemOpen
  \bibfield  {author} {\bibinfo {author} {\bibfnamefont {J.}~\bibnamefont
  {Zhang}}, \bibinfo {author} {\bibfnamefont {Y.}~\bibnamefont {Yao}}, \ and\
  \bibinfo {author} {\bibfnamefont {W.}~\bibnamefont {Kong}},\ }\href {\doibase
  10.1021/acs.jpclett.0c00092} {\bibfield  {journal} {\bibinfo  {journal} {J.
  Phys. Chem. Lett.}\ }\textbf {\bibinfo {volume} {11}},\ \bibinfo {pages}
  {1100} (\bibinfo {year} {2020})}\BibitemShut {NoStop}%
\bibitem [{\citenamefont {Sch\"utte}\ \emph {et~al.}(2018)\citenamefont
  {Sch\"utte}, \citenamefont {Peltz}, \citenamefont {Austin}, \citenamefont
  {Str\"uber}, \citenamefont {Ye}, \citenamefont {Rouz\'ee}, \citenamefont
  {Vrakking}, \citenamefont {Golubev}, \citenamefont {Kuleff}, \citenamefont
  {Fennel},\ and\ \citenamefont {Marangos}}]{SchPRL18}%
  \BibitemOpen
  \bibfield  {author} {\bibinfo {author} {\bibfnamefont {B.}~\bibnamefont
  {Sch\"utte}}, \bibinfo {author} {\bibfnamefont {C.}~\bibnamefont {Peltz}},
  \bibinfo {author} {\bibfnamefont {D.~R.}\ \bibnamefont {Austin}}, \bibinfo
  {author} {\bibfnamefont {C.}~\bibnamefont {Str\"uber}}, \bibinfo {author}
  {\bibfnamefont {P.}~\bibnamefont {Ye}}, \bibinfo {author} {\bibfnamefont
  {A.}~\bibnamefont {Rouz\'ee}}, \bibinfo {author} {\bibfnamefont {M.~J.~J.}\
  \bibnamefont {Vrakking}}, \bibinfo {author} {\bibfnamefont {N.}~\bibnamefont
  {Golubev}}, \bibinfo {author} {\bibfnamefont {A.~I.}\ \bibnamefont {Kuleff}},
  \bibinfo {author} {\bibfnamefont {T.}~\bibnamefont {Fennel}}, \ and\ \bibinfo
  {author} {\bibfnamefont {J.~P.}\ \bibnamefont {Marangos}},\ }\href {\doibase
  10.1103/PhysRevLett.121.063202} {\bibfield  {journal} {\bibinfo  {journal}
  {Phys. Rev. Lett.}\ }\textbf {\bibinfo {volume} {121}},\ \bibinfo {pages}
  {063202} (\bibinfo {year} {2018})}\BibitemShut {NoStop}%
\bibitem [{\citenamefont {Streeter}\ \emph {et~al.}(2018)\citenamefont
  {Streeter}, \citenamefont {Dann}, \citenamefont {Scott}, \citenamefont
  {Baird}, \citenamefont {Murphy}, \citenamefont {Eardley}, \citenamefont
  {Smith}, \citenamefont {Rozario}, \citenamefont {Gruse}, \citenamefont
  {Mangles}, \citenamefont {Najmudin}, \citenamefont {Tata}, \citenamefont
  {Krishnamurthy}, \citenamefont {Rahul}, \citenamefont {Hazra}, \citenamefont
  {Pourmoussavi}, \citenamefont {Osterhoff}, \citenamefont {Hah}, \citenamefont
  {Bourgeois}, \citenamefont {Thornton}, \citenamefont {Gregory}, \citenamefont
  {Hooker}, \citenamefont {Chekhlov}, \citenamefont {Hawkes}, \citenamefont
  {Parry}, \citenamefont {Marshall}, \citenamefont {Tang}, \citenamefont
  {Springate}, \citenamefont {Rajeev}, \citenamefont {Thomas},\ and\
  \citenamefont {Symes}}]{StrAPL18}%
  \BibitemOpen
  \bibfield  {author} {\bibinfo {author} {\bibfnamefont {M.~J.~V.}\
  \bibnamefont {Streeter}}, \bibinfo {author} {\bibfnamefont {S.~J.~D.}\
  \bibnamefont {Dann}}, \bibinfo {author} {\bibfnamefont {J.~D.~E.}\
  \bibnamefont {Scott}}, \bibinfo {author} {\bibfnamefont {C.~D.}\ \bibnamefont
  {Baird}}, \bibinfo {author} {\bibfnamefont {C.~D.}\ \bibnamefont {Murphy}},
  \bibinfo {author} {\bibfnamefont {S.}~\bibnamefont {Eardley}}, \bibinfo
  {author} {\bibfnamefont {R.~A.}\ \bibnamefont {Smith}}, \bibinfo {author}
  {\bibfnamefont {S.}~\bibnamefont {Rozario}}, \bibinfo {author} {\bibfnamefont
  {J.-N.}\ \bibnamefont {Gruse}}, \bibinfo {author} {\bibfnamefont {S.~P.~D.}\
  \bibnamefont {Mangles}}, \bibinfo {author} {\bibfnamefont {Z.}~\bibnamefont
  {Najmudin}}, \bibinfo {author} {\bibfnamefont {S.}~\bibnamefont {Tata}},
  \bibinfo {author} {\bibfnamefont {M.}~\bibnamefont {Krishnamurthy}}, \bibinfo
  {author} {\bibfnamefont {S.~V.}\ \bibnamefont {Rahul}}, \bibinfo {author}
  {\bibfnamefont {D.}~\bibnamefont {Hazra}}, \bibinfo {author} {\bibfnamefont
  {P.}~\bibnamefont {Pourmoussavi}}, \bibinfo {author} {\bibfnamefont
  {J.}~\bibnamefont {Osterhoff}}, \bibinfo {author} {\bibfnamefont
  {J.}~\bibnamefont {Hah}}, \bibinfo {author} {\bibfnamefont {N.}~\bibnamefont
  {Bourgeois}}, \bibinfo {author} {\bibfnamefont {C.}~\bibnamefont {Thornton}},
  \bibinfo {author} {\bibfnamefont {C.~D.}\ \bibnamefont {Gregory}}, \bibinfo
  {author} {\bibfnamefont {C.~J.}\ \bibnamefont {Hooker}}, \bibinfo {author}
  {\bibfnamefont {O.}~\bibnamefont {Chekhlov}}, \bibinfo {author}
  {\bibfnamefont {S.~J.}\ \bibnamefont {Hawkes}}, \bibinfo {author}
  {\bibfnamefont {B.}~\bibnamefont {Parry}}, \bibinfo {author} {\bibfnamefont
  {V.~A.}\ \bibnamefont {Marshall}}, \bibinfo {author} {\bibfnamefont
  {Y.}~\bibnamefont {Tang}}, \bibinfo {author} {\bibfnamefont {E.}~\bibnamefont
  {Springate}}, \bibinfo {author} {\bibfnamefont {P.~P.}\ \bibnamefont
  {Rajeev}}, \bibinfo {author} {\bibfnamefont {A.~G.~R.}\ \bibnamefont
  {Thomas}}, \ and\ \bibinfo {author} {\bibfnamefont {D.~R.}\ \bibnamefont
  {Symes}},\ }\href {\doibase 10.1063/1.5027297} {\bibfield  {journal}
  {\bibinfo  {journal} {Appl. Phys. Lett.}\ }\textbf {\bibinfo {volume}
  {112}},\ \bibinfo {pages} {244101} (\bibinfo {year} {2018})}\BibitemShut
  {NoStop}%
\bibitem [{\citenamefont {Augst}\ \emph {et~al.}(1989)\citenamefont {Augst},
  \citenamefont {Strickland}, \citenamefont {Meyerhofer}, \citenamefont
  {Chin},\ and\ \citenamefont {Eberly}}]{AugPRL89}%
  \BibitemOpen
  \bibfield  {author} {\bibinfo {author} {\bibfnamefont {S.}~\bibnamefont
  {Augst}}, \bibinfo {author} {\bibfnamefont {D.}~\bibnamefont {Strickland}},
  \bibinfo {author} {\bibfnamefont {D.~D.}\ \bibnamefont {Meyerhofer}},
  \bibinfo {author} {\bibfnamefont {S.~L.}\ \bibnamefont {Chin}}, \ and\
  \bibinfo {author} {\bibfnamefont {J.~H.}\ \bibnamefont {Eberly}},\
  }\href@noop {} {\bibfield  {journal} {\bibinfo  {journal} {Phys. Rev. Lett.}\
  }\textbf {\bibinfo {volume} {63}},\ \bibinfo {pages} {2212} (\bibinfo {year}
  {1989})}\BibitemShut {NoStop}%
\bibitem [{\citenamefont {Truong}\ \emph {et~al.}(2011)\citenamefont {Truong},
  \citenamefont {G\"ode}, \citenamefont {Tiggesb\"aumker},\ and\ \citenamefont
  {Meiwes-Broer}}]{TruEPJD11}%
  \BibitemOpen
  \bibfield  {author} {\bibinfo {author} {\bibfnamefont {N.}~\bibnamefont
  {Truong}}, \bibinfo {author} {\bibfnamefont {S.}~\bibnamefont {G\"ode}},
  \bibinfo {author} {\bibfnamefont {J.}~\bibnamefont {Tiggesb\"aumker}}, \ and\
  \bibinfo {author} {\bibfnamefont {K.-H.}\ \bibnamefont {Meiwes-Broer}},\
  }\href {\doibase 10.1140/epjd/e2011-10533-6} {\bibfield  {journal} {\bibinfo
  {journal} {Eur. Phys. J. D}\ }\textbf {\bibinfo {volume} {63}},\ \bibinfo
  {pages} {275} (\bibinfo {year} {2011})}\BibitemShut {NoStop}%
\bibitem [{\citenamefont {Islam}\ \emph {et~al.}(2006)\citenamefont {Islam},
  \citenamefont {Saalmann},\ and\ \citenamefont {Rost}}]{IslPRA06}%
  \BibitemOpen
  \bibfield  {author} {\bibinfo {author} {\bibfnamefont {M.}~\bibnamefont
  {Islam}}, \bibinfo {author} {\bibfnamefont {U.}~\bibnamefont {Saalmann}}, \
  and\ \bibinfo {author} {\bibfnamefont {J.~M.}\ \bibnamefont {Rost}},\
  }\href@noop {} {\bibfield  {journal} {\bibinfo  {journal} {Phys. Rev. A}\
  }\textbf {\bibinfo {volume} {73}},\ \bibinfo {pages} {041201} (\bibinfo
  {year} {2006})}\BibitemShut {NoStop}%
\bibitem [{\citenamefont {Rose-Petruck}\ \emph {et~al.}(1997)\citenamefont
  {Rose-Petruck}, \citenamefont {Sch\"afer}, \citenamefont {Wilson},\ and\
  \citenamefont {Barty}}]{RosPRA97}%
  \BibitemOpen
  \bibfield  {author} {\bibinfo {author} {\bibfnamefont {C.}~\bibnamefont
  {Rose-Petruck}}, \bibinfo {author} {\bibfnamefont {K.}~\bibnamefont
  {Sch\"afer}}, \bibinfo {author} {\bibfnamefont {K.}~\bibnamefont {Wilson}}, \
  and\ \bibinfo {author} {\bibfnamefont {C.}~\bibnamefont {Barty}},\
  }\href@noop {} {\bibfield  {journal} {\bibinfo  {journal} {Phys. Rev. A}\
  }\textbf {\bibinfo {volume} {55}},\ \bibinfo {pages} {1182} (\bibinfo {year}
  {1997})}\BibitemShut {NoStop}%
\bibitem [{\citenamefont {Ditmire}\ \emph {et~al.}(1998)\citenamefont
  {Ditmire}, \citenamefont {Springate}, \citenamefont {Tisch}, \citenamefont
  {Shao}, \citenamefont {Mason}, \citenamefont {Hay}, \citenamefont
  {Marangos},\ and\ \citenamefont {Hutchinson}}]{DitPRA98}%
  \BibitemOpen
  \bibfield  {author} {\bibinfo {author} {\bibfnamefont {T.}~\bibnamefont
  {Ditmire}}, \bibinfo {author} {\bibfnamefont {E.}~\bibnamefont {Springate}},
  \bibinfo {author} {\bibfnamefont {J.}~\bibnamefont {Tisch}}, \bibinfo
  {author} {\bibfnamefont {Y.}~\bibnamefont {Shao}}, \bibinfo {author}
  {\bibfnamefont {M.}~\bibnamefont {Mason}}, \bibinfo {author} {\bibfnamefont
  {N.}~\bibnamefont {Hay}}, \bibinfo {author} {\bibfnamefont {J.}~\bibnamefont
  {Marangos}}, \ and\ \bibinfo {author} {\bibfnamefont {M.}~\bibnamefont
  {Hutchinson}},\ }\href@noop {} {\bibfield  {journal} {\bibinfo  {journal}
  {Phys. Rev. A}\ }\textbf {\bibinfo {volume} {57}},\ \bibinfo {pages} {369}
  (\bibinfo {year} {1998})}\BibitemShut {NoStop}%
\bibitem [{\citenamefont {Last}\ and\ \citenamefont
  {Jortner}(1998)}]{LasPRA98}%
  \BibitemOpen
  \bibfield  {author} {\bibinfo {author} {\bibfnamefont {I.}~\bibnamefont
  {Last}}\ and\ \bibinfo {author} {\bibfnamefont {J.}~\bibnamefont {Jortner}},\
  }\href@noop {} {\bibfield  {journal} {\bibinfo  {journal} {Phys. Rev. A}\
  }\textbf {\bibinfo {volume} {58}},\ \bibinfo {pages} {3826} (\bibinfo {year}
  {1998})}\BibitemShut {NoStop}%
\bibitem [{\citenamefont {Ishikawa}\ and\ \citenamefont
  {Blenski}(2000)}]{IshPRA00}%
  \BibitemOpen
  \bibfield  {author} {\bibinfo {author} {\bibfnamefont {K.}~\bibnamefont
  {Ishikawa}}\ and\ \bibinfo {author} {\bibfnamefont {T.}~\bibnamefont
  {Blenski}},\ }\href@noop {} {\bibfield  {journal} {\bibinfo  {journal} {Phys.
  Rev. A}\ }\textbf {\bibinfo {volume} {62}},\ \bibinfo {pages} {63204}
  (\bibinfo {year} {2000})}\BibitemShut {NoStop}%
\bibitem [{\citenamefont {Siedschlag}\ and\ \citenamefont
  {Rost}(2002)}]{SiePRL02}%
  \BibitemOpen
  \bibfield  {author} {\bibinfo {author} {\bibfnamefont {C.}~\bibnamefont
  {Siedschlag}}\ and\ \bibinfo {author} {\bibfnamefont {J.}~\bibnamefont
  {Rost}},\ }\href@noop {} {\bibfield  {journal} {\bibinfo  {journal} {Phys.
  Rev. Lett.}\ }\textbf {\bibinfo {volume} {89}},\ \bibinfo {pages} {173401}
  (\bibinfo {year} {2002})}\BibitemShut {NoStop}%
\bibitem [{\citenamefont {Bauer}(2004)}]{BauJPB04}%
  \BibitemOpen
  \bibfield  {author} {\bibinfo {author} {\bibfnamefont {D.}~\bibnamefont
  {Bauer}},\ }\href@noop {} {\bibfield  {journal} {\bibinfo  {journal} {J.
  Phys. B}\ }\textbf {\bibinfo {volume} {37}},\ \bibinfo {pages} {3085}
  (\bibinfo {year} {2004})}\BibitemShut {NoStop}%
\bibitem [{\citenamefont {Fennel}\ \emph
  {et~al.}(2007{\natexlab{b}})\citenamefont {Fennel}, \citenamefont {Ramunno},\
  and\ \citenamefont {Brabec}}]{FenPRL07b}%
  \BibitemOpen
  \bibfield  {author} {\bibinfo {author} {\bibfnamefont {T.}~\bibnamefont
  {Fennel}}, \bibinfo {author} {\bibfnamefont {L.}~\bibnamefont {Ramunno}}, \
  and\ \bibinfo {author} {\bibfnamefont {T.}~\bibnamefont {Brabec}},\
  }\href@noop {} {\bibfield  {journal} {\bibinfo  {journal} {Phys. Rev. Lett.}\
  }\textbf {\bibinfo {volume} {99}},\ \bibinfo {pages} {233401} (\bibinfo
  {year} {2007}{\natexlab{b}})}\BibitemShut {NoStop}%
\bibitem [{\citenamefont {Petrov}\ and\ \citenamefont {Davis}(2008)}]{PetPP08}%
  \BibitemOpen
  \bibfield  {author} {\bibinfo {author} {\bibfnamefont {G.~M.}\ \bibnamefont
  {Petrov}}\ and\ \bibinfo {author} {\bibfnamefont {J.}~\bibnamefont {Davis}},\
  }\href@noop {} {\bibfield  {journal} {\bibinfo  {journal} {Phys. Plas.}\
  }\textbf {\bibinfo {volume} {15}},\ \bibinfo {pages} {056705} (\bibinfo
  {year} {2008})}\BibitemShut {NoStop}%
\bibitem [{\citenamefont {Ditmire}\ \emph {et~al.}(1997)\citenamefont
  {Ditmire}, \citenamefont {Tisch}, \citenamefont {Springate}, \citenamefont
  {Mason}, \citenamefont {Hay}, \citenamefont {Smith}, \citenamefont
  {Marangos},\ and\ \citenamefont {Hutchinson}}]{DitN97}%
  \BibitemOpen
  \bibfield  {author} {\bibinfo {author} {\bibfnamefont {T.}~\bibnamefont
  {Ditmire}}, \bibinfo {author} {\bibfnamefont {J.}~\bibnamefont {Tisch}},
  \bibinfo {author} {\bibfnamefont {E.}~\bibnamefont {Springate}}, \bibinfo
  {author} {\bibfnamefont {M.}~\bibnamefont {Mason}}, \bibinfo {author}
  {\bibfnamefont {N.}~\bibnamefont {Hay}}, \bibinfo {author} {\bibfnamefont
  {R.}~\bibnamefont {Smith}}, \bibinfo {author} {\bibfnamefont
  {I.}~\bibnamefont {Marangos}}, \ and\ \bibinfo {author} {\bibfnamefont
  {M.}~\bibnamefont {Hutchinson}},\ }\href@noop {} {\bibfield  {journal}
  {\bibinfo  {journal} {Nature}\ }\textbf {\bibinfo {volume} {386}},\ \bibinfo
  {pages} {54} (\bibinfo {year} {1997})}\BibitemShut {NoStop}%
\bibitem [{\citenamefont {Kumarappan}\ \emph {et~al.}(2001)\citenamefont
  {Kumarappan}, \citenamefont {Krishnamurthy},\ and\ \citenamefont
  {Mathur}}]{KumPRL01}%
  \BibitemOpen
  \bibfield  {author} {\bibinfo {author} {\bibfnamefont {V.}~\bibnamefont
  {Kumarappan}}, \bibinfo {author} {\bibfnamefont {M.}~\bibnamefont
  {Krishnamurthy}}, \ and\ \bibinfo {author} {\bibfnamefont {D.}~\bibnamefont
  {Mathur}},\ }\href@noop {} {\bibfield  {journal} {\bibinfo  {journal} {Phys.
  Rev. Lett.}\ }\textbf {\bibinfo {volume} {87}},\ \bibinfo {pages} {085005}
  (\bibinfo {year} {2001})}\BibitemShut {NoStop}%
\bibitem [{\citenamefont {Sakabe}\ \emph {et~al.}(2004)\citenamefont {Sakabe},
  \citenamefont {Shimizu}, \citenamefont {Hashida}, \citenamefont {Sato},
  \citenamefont {Tsuyukushi}, \citenamefont {Nishihara}, \citenamefont
  {Okihara}, \citenamefont {Kagawa}, \citenamefont {Izawa}, \citenamefont
  {Imasaki},\ and\ \citenamefont {Iida}}]{SakPRA04}%
  \BibitemOpen
  \bibfield  {author} {\bibinfo {author} {\bibfnamefont {S.}~\bibnamefont
  {Sakabe}}, \bibinfo {author} {\bibfnamefont {S.}~\bibnamefont {Shimizu}},
  \bibinfo {author} {\bibfnamefont {M.}~\bibnamefont {Hashida}}, \bibinfo
  {author} {\bibfnamefont {F.}~\bibnamefont {Sato}}, \bibinfo {author}
  {\bibfnamefont {T.}~\bibnamefont {Tsuyukushi}}, \bibinfo {author}
  {\bibfnamefont {K.}~\bibnamefont {Nishihara}}, \bibinfo {author}
  {\bibfnamefont {S.}~\bibnamefont {Okihara}}, \bibinfo {author} {\bibfnamefont
  {T.}~\bibnamefont {Kagawa}}, \bibinfo {author} {\bibfnamefont
  {Y.}~\bibnamefont {Izawa}}, \bibinfo {author} {\bibfnamefont
  {K.}~\bibnamefont {Imasaki}}, \ and\ \bibinfo {author} {\bibfnamefont
  {T.}~\bibnamefont {Iida}},\ }\href@noop {} {\bibfield  {journal} {\bibinfo
  {journal} {Phys. Rev. A}\ }\textbf {\bibinfo {volume} {69}},\ \bibinfo
  {pages} {023203} (\bibinfo {year} {2004})}\BibitemShut {NoStop}%
\bibitem [{\citenamefont {Krishnamurthy}\ \emph {et~al.}(2004)\citenamefont
  {Krishnamurthy}, \citenamefont {Mathur},\ and\ \citenamefont
  {Kumarappan}}]{KriPRA04}%
  \BibitemOpen
  \bibfield  {author} {\bibinfo {author} {\bibfnamefont {M.}~\bibnamefont
  {Krishnamurthy}}, \bibinfo {author} {\bibfnamefont {D.}~\bibnamefont
  {Mathur}}, \ and\ \bibinfo {author} {\bibfnamefont {V.}~\bibnamefont
  {Kumarappan}},\ }\href@noop {} {\bibfield  {journal} {\bibinfo  {journal}
  {Phys. Rev. A}\ }\textbf {\bibinfo {volume} {69}},\ \bibinfo {pages} {033202}
  (\bibinfo {year} {2004})}\BibitemShut {NoStop}%
\bibitem [{\citenamefont {Rajeev}\ \emph {et~al.}(2011)\citenamefont {Rajeev},
  \citenamefont {Rishad}, \citenamefont {Trivikram}, \citenamefont
  {Narayanan},\ and\ \citenamefont {Krishnamurthy}}]{RajRSI11}%
  \BibitemOpen
  \bibfield  {author} {\bibinfo {author} {\bibfnamefont {R.}~\bibnamefont
  {Rajeev}}, \bibinfo {author} {\bibfnamefont {K.~P.~M.}\ \bibnamefont
  {Rishad}}, \bibinfo {author} {\bibfnamefont {T.~M.}\ \bibnamefont
  {Trivikram}}, \bibinfo {author} {\bibfnamefont {V.}~\bibnamefont
  {Narayanan}}, \ and\ \bibinfo {author} {\bibfnamefont {M.}~\bibnamefont
  {Krishnamurthy}},\ }\href {\doibase 10.1063/1.3624698} {\bibfield  {journal}
  {\bibinfo  {journal} {Rev. Sci. Instr.}\ }\textbf {\bibinfo {volume} {82}},\
  \bibinfo {pages} {083303} (\bibinfo {year} {2011})}\BibitemShut {NoStop}%
\bibitem [{\citenamefont {Iwayama}\ \emph {et~al.}(2013)\citenamefont
  {Iwayama}, \citenamefont {Nagaya}, \citenamefont {Yao}, \citenamefont
  {Fukuzawa}, \citenamefont {Liu}, \citenamefont {Pr\"umper}, \citenamefont
  {Motomura}, \citenamefont {Ueda}, \citenamefont {Saito}, \citenamefont
  {Rudenko}, \citenamefont {Foucar}, \citenamefont {Nagasono}, \citenamefont
  {Higashiya}, \citenamefont {Yabashi}, \citenamefont {Ishikawa}, \citenamefont
  {Ohashi},\ and\ \citenamefont {Kimura}}]{IwaJPB13}%
  \BibitemOpen
  \bibfield  {author} {\bibinfo {author} {\bibfnamefont {H.}~\bibnamefont
  {Iwayama}}, \bibinfo {author} {\bibfnamefont {K.}~\bibnamefont {Nagaya}},
  \bibinfo {author} {\bibfnamefont {M.}~\bibnamefont {Yao}}, \bibinfo {author}
  {\bibfnamefont {H.}~\bibnamefont {Fukuzawa}}, \bibinfo {author}
  {\bibfnamefont {X.-J.}\ \bibnamefont {Liu}}, \bibinfo {author} {\bibfnamefont
  {G.}~\bibnamefont {Pr\"umper}}, \bibinfo {author} {\bibfnamefont
  {K.}~\bibnamefont {Motomura}}, \bibinfo {author} {\bibfnamefont
  {K.}~\bibnamefont {Ueda}}, \bibinfo {author} {\bibfnamefont {N.}~\bibnamefont
  {Saito}}, \bibinfo {author} {\bibfnamefont {A.}~\bibnamefont {Rudenko}},
  \bibinfo {author} {\bibfnamefont {L.}~\bibnamefont {Foucar}}, \bibinfo
  {author} {\bibfnamefont {M.}~\bibnamefont {Nagasono}}, \bibinfo {author}
  {\bibfnamefont {A.}~\bibnamefont {Higashiya}}, \bibinfo {author}
  {\bibfnamefont {M.}~\bibnamefont {Yabashi}}, \bibinfo {author} {\bibfnamefont
  {T.}~\bibnamefont {Ishikawa}}, \bibinfo {author} {\bibfnamefont
  {H.}~\bibnamefont {Ohashi}}, \ and\ \bibinfo {author} {\bibfnamefont
  {H.}~\bibnamefont {Kimura}},\ }\href {\doibase
  10.1088/0953-4075/46/16/164019} {\bibfield  {journal} {\bibinfo  {journal}
  {J. Phys. B}\ }\textbf {\bibinfo {volume} {46}},\ \bibinfo {pages} {164019}
  (\bibinfo {year} {2013})}\BibitemShut {NoStop}%
\bibitem [{\citenamefont {Komar}\ \emph {et~al.}(2016)\citenamefont {Komar},
  \citenamefont {Meiwes-Broer},\ and\ \citenamefont
  {Tiggesb\"aumker}}]{KomRSI16}%
  \BibitemOpen
  \bibfield  {author} {\bibinfo {author} {\bibfnamefont {D.}~\bibnamefont
  {Komar}}, \bibinfo {author} {\bibfnamefont {K.-H.}\ \bibnamefont
  {Meiwes-Broer}}, \ and\ \bibinfo {author} {\bibfnamefont {J.}~\bibnamefont
  {Tiggesb\"aumker}},\ }\href {\doibase 10.1063/1.4964474} {\bibfield
  {journal} {\bibinfo  {journal} {Rev. Sci. Instr.}\ }\textbf {\bibinfo
  {volume} {87}},\ \bibinfo {pages} {103110} (\bibinfo {year}
  {2016})}\BibitemShut {NoStop}%
\bibitem [{\citenamefont {Pentlehner}\ \emph {et~al.}(2009)\citenamefont
  {Pentlehner}, \citenamefont {Riechers}, \citenamefont {Dick}, \citenamefont
  {Slenczka}, \citenamefont {Even}, \citenamefont {Lavie}, \citenamefont
  {Brown},\ and\ \citenamefont {Luria}}]{PenRSI09}%
  \BibitemOpen
  \bibfield  {author} {\bibinfo {author} {\bibfnamefont {D.}~\bibnamefont
  {Pentlehner}}, \bibinfo {author} {\bibfnamefont {R.}~\bibnamefont
  {Riechers}}, \bibinfo {author} {\bibfnamefont {B.}~\bibnamefont {Dick}},
  \bibinfo {author} {\bibfnamefont {A.}~\bibnamefont {Slenczka}}, \bibinfo
  {author} {\bibfnamefont {U.}~\bibnamefont {Even}}, \bibinfo {author}
  {\bibfnamefont {N.}~\bibnamefont {Lavie}}, \bibinfo {author} {\bibfnamefont
  {R.}~\bibnamefont {Brown}}, \ and\ \bibinfo {author} {\bibfnamefont
  {K.}~\bibnamefont {Luria}},\ }\href {\doibase 10.1063/1.3117196} {\bibfield
  {journal} {\bibinfo  {journal} {Rev. Sci. Instr.}\ }\textbf {\bibinfo
  {volume} {80}},\ \bibinfo {pages} {043302} (\bibinfo {year}
  {2009})}\BibitemShut {NoStop}%
\bibitem [{\citenamefont {K\"oller}\ \emph {et~al.}(1999)\citenamefont
  {K\"oller}, \citenamefont {Schumacher}, \citenamefont {K\"ohn}, \citenamefont
  {Teuber}, \citenamefont {Tiggesb\"aumker},\ and\ \citenamefont
  {Meiwes-Broer}}]{KoePRL99}%
  \BibitemOpen
  \bibfield  {author} {\bibinfo {author} {\bibfnamefont {L.}~\bibnamefont
  {K\"oller}}, \bibinfo {author} {\bibfnamefont {M.}~\bibnamefont
  {Schumacher}}, \bibinfo {author} {\bibfnamefont {J.}~\bibnamefont {K\"ohn}},
  \bibinfo {author} {\bibfnamefont {S.}~\bibnamefont {Teuber}}, \bibinfo
  {author} {\bibfnamefont {J.}~\bibnamefont {Tiggesb\"aumker}}, \ and\ \bibinfo
  {author} {\bibfnamefont {K.-H.}\ \bibnamefont {Meiwes-Broer}},\ }\href@noop
  {} {\bibfield  {journal} {\bibinfo  {journal} {Phys. Rev. Lett.}\ }\textbf
  {\bibinfo {volume} {82}},\ \bibinfo {pages} {3783} (\bibinfo {year}
  {1999})}\BibitemShut {NoStop}%
\bibitem [{\citenamefont {Last}\ and\ \citenamefont
  {Jortner}(2004)}]{LasJCP04a}%
  \BibitemOpen
  \bibfield  {author} {\bibinfo {author} {\bibfnamefont {I.}~\bibnamefont
  {Last}}\ and\ \bibinfo {author} {\bibfnamefont {J.}~\bibnamefont {Jortner}},\
  }\href@noop {} {\bibfield  {journal} {\bibinfo  {journal} {J. Chem. Phys.}\
  }\textbf {\bibinfo {volume} {120}},\ \bibinfo {pages} {1336} (\bibinfo {year}
  {2004})}\BibitemShut {NoStop}%
\bibitem [{\citenamefont {Rajeev}\ \emph {et~al.}(2012)\citenamefont {Rajeev},
  \citenamefont {Rishad}, \citenamefont {Trivikram}, \citenamefont {Narayanan},
  \citenamefont {Brabec},\ and\ \citenamefont {Krishnamurthy}}]{RajPRA12}%
  \BibitemOpen
  \bibfield  {author} {\bibinfo {author} {\bibfnamefont {R.}~\bibnamefont
  {Rajeev}}, \bibinfo {author} {\bibfnamefont {K.~P.~M.}\ \bibnamefont
  {Rishad}}, \bibinfo {author} {\bibfnamefont {T.~M.}\ \bibnamefont
  {Trivikram}}, \bibinfo {author} {\bibfnamefont {V.}~\bibnamefont
  {Narayanan}}, \bibinfo {author} {\bibfnamefont {T.}~\bibnamefont {Brabec}}, \
  and\ \bibinfo {author} {\bibfnamefont {M.}~\bibnamefont {Krishnamurthy}},\
  }\href {\doibase 10.1103/PhysRevA.85.023201} {\bibfield  {journal} {\bibinfo
  {journal} {Phys. Rev. A}\ }\textbf {\bibinfo {volume} {85}},\ \bibinfo
  {pages} {023201} (\bibinfo {year} {2012})}\BibitemShut {NoStop}%
\bibitem [{\citenamefont {Bethe}\ and\ \citenamefont {Salpeter}(1977)}]{Bet77}%
  \BibitemOpen
  \bibfield  {author} {\bibinfo {author} {\bibfnamefont {H.}~\bibnamefont
  {Bethe}}\ and\ \bibinfo {author} {\bibfnamefont {E.}~\bibnamefont
  {Salpeter}},\ }\href@noop {} {\emph {\bibinfo {title} {{Quantum Mechanics of
  One- and Two-Electron Atoms}}}}\ (\bibinfo  {publisher} {Plenum Press},\
  \bibinfo {address} {New York},\ \bibinfo {year} {1977})\BibitemShut {NoStop}%
\bibitem [{\citenamefont {T.~D\"oppner~and}\ \emph {et~al.}(2007)\citenamefont
  {T.~D\"oppner~and}, \citenamefont {Przystawik}, \citenamefont
  {Tiggesb\"aumker},\ and\ \citenamefont {Meiwes-Broer}}]{DoeEPJD07}%
  \BibitemOpen
  \bibfield  {author} {\bibinfo {author} {\bibfnamefont {J.~P.~M.}\
  \bibnamefont {T.~D\"oppner~and}}, \bibinfo {author} {\bibfnamefont
  {A.}~\bibnamefont {Przystawik}}, \bibinfo {author} {\bibfnamefont
  {J.}~\bibnamefont {Tiggesb\"aumker}}, \ and\ \bibinfo {author} {\bibfnamefont
  {K.-H.}\ \bibnamefont {Meiwes-Broer}},\ }\href@noop {} {\bibfield  {journal}
  {\bibinfo  {journal} {Eur. Phys. J. D}\ }\textbf {\bibinfo {volume} {43}},\
  \bibinfo {pages} {261} (\bibinfo {year} {2007})}\BibitemShut {NoStop}%
\bibitem [{\citenamefont {Kumarappan}\ \emph {et~al.}(2002)\citenamefont
  {Kumarappan}, \citenamefont {Krishnamurthy},\ and\ \citenamefont
  {Mathur}}]{KumPRA02}%
  \BibitemOpen
  \bibfield  {author} {\bibinfo {author} {\bibfnamefont {V.}~\bibnamefont
  {Kumarappan}}, \bibinfo {author} {\bibfnamefont {M.}~\bibnamefont
  {Krishnamurthy}}, \ and\ \bibinfo {author} {\bibfnamefont {D.}~\bibnamefont
  {Mathur}},\ }\href@noop {} {\bibfield  {journal} {\bibinfo  {journal} {Phys.
  Rev. A}\ }\textbf {\bibinfo {volume} {66}},\ \bibinfo {pages} {33203}
  (\bibinfo {year} {2002})}\BibitemShut {NoStop}%
\bibitem [{\citenamefont {Springate}\ \emph {et~al.}(2000)\citenamefont
  {Springate}, \citenamefont {Hay}, \citenamefont {Tisch}, \citenamefont
  {Mason}, \citenamefont {Ditmire}, \citenamefont {Marangos},\ and\
  \citenamefont {Hutchinson}}]{SprPRA00}%
  \BibitemOpen
  \bibfield  {author} {\bibinfo {author} {\bibfnamefont {E.}~\bibnamefont
  {Springate}}, \bibinfo {author} {\bibfnamefont {N.}~\bibnamefont {Hay}},
  \bibinfo {author} {\bibfnamefont {J.~W.~G.}\ \bibnamefont {Tisch}}, \bibinfo
  {author} {\bibfnamefont {M.~B.}\ \bibnamefont {Mason}}, \bibinfo {author}
  {\bibfnamefont {T.}~\bibnamefont {Ditmire}}, \bibinfo {author} {\bibfnamefont
  {J.~P.}\ \bibnamefont {Marangos}}, \ and\ \bibinfo {author} {\bibfnamefont
  {M.~H.~R.}\ \bibnamefont {Hutchinson}},\ }\href@noop {} {\bibfield  {journal}
  {\bibinfo  {journal} {Phys. Rev. A}\ }\textbf {\bibinfo {volume} {61}},\
  \bibinfo {pages} {44101} (\bibinfo {year} {2000})}\BibitemShut {NoStop}%
\bibitem [{\citenamefont {Thomas}\ \emph {et~al.}(2012)\citenamefont {Thomas},
  \citenamefont {Helal}, \citenamefont {Hoffmann}, \citenamefont {Kandadai},
  \citenamefont {Keto}, \citenamefont {Andreasson}, \citenamefont {Iwan},
  \citenamefont {Seibert}, \citenamefont {Timneanu}, \citenamefont {Hajdu},
  \citenamefont {Adolph}, \citenamefont {Gorkhover}, \citenamefont {Rupp},
  \citenamefont {Schorb}, \citenamefont {M\"oller}, \citenamefont {Doumy},
  \citenamefont {DiMauro}, \citenamefont {Hoener}, \citenamefont {Murphy},
  \citenamefont {Berrah}, \citenamefont {Messerschmidt}, \citenamefont {Bozek},
  \citenamefont {Bostedt},\ and\ \citenamefont {Ditmire}}]{ThoPRL12}%
  \BibitemOpen
  \bibfield  {author} {\bibinfo {author} {\bibfnamefont {H.}~\bibnamefont
  {Thomas}}, \bibinfo {author} {\bibfnamefont {A.}~\bibnamefont {Helal}},
  \bibinfo {author} {\bibfnamefont {K.}~\bibnamefont {Hoffmann}}, \bibinfo
  {author} {\bibfnamefont {N.}~\bibnamefont {Kandadai}}, \bibinfo {author}
  {\bibfnamefont {J.}~\bibnamefont {Keto}}, \bibinfo {author} {\bibfnamefont
  {J.}~\bibnamefont {Andreasson}}, \bibinfo {author} {\bibfnamefont
  {B.}~\bibnamefont {Iwan}}, \bibinfo {author} {\bibfnamefont {M.}~\bibnamefont
  {Seibert}}, \bibinfo {author} {\bibfnamefont {N.}~\bibnamefont {Timneanu}},
  \bibinfo {author} {\bibfnamefont {J.}~\bibnamefont {Hajdu}}, \bibinfo
  {author} {\bibfnamefont {M.}~\bibnamefont {Adolph}}, \bibinfo {author}
  {\bibfnamefont {T.}~\bibnamefont {Gorkhover}}, \bibinfo {author}
  {\bibfnamefont {D.}~\bibnamefont {Rupp}}, \bibinfo {author} {\bibfnamefont
  {S.}~\bibnamefont {Schorb}}, \bibinfo {author} {\bibfnamefont
  {T.}~\bibnamefont {M\"oller}}, \bibinfo {author} {\bibfnamefont
  {G.}~\bibnamefont {Doumy}}, \bibinfo {author} {\bibfnamefont {L.~F.}\
  \bibnamefont {DiMauro}}, \bibinfo {author} {\bibfnamefont {M.}~\bibnamefont
  {Hoener}}, \bibinfo {author} {\bibfnamefont {B.}~\bibnamefont {Murphy}},
  \bibinfo {author} {\bibfnamefont {N.}~\bibnamefont {Berrah}}, \bibinfo
  {author} {\bibfnamefont {M.}~\bibnamefont {Messerschmidt}}, \bibinfo {author}
  {\bibfnamefont {J.}~\bibnamefont {Bozek}}, \bibinfo {author} {\bibfnamefont
  {C.}~\bibnamefont {Bostedt}}, \ and\ \bibinfo {author} {\bibfnamefont
  {T.}~\bibnamefont {Ditmire}},\ }\href {\doibase
  10.1103/PhysRevLett.108.133401} {\bibfield  {journal} {\bibinfo  {journal}
  {Phys. Rev. Lett.}\ }\textbf {\bibinfo {volume} {108}},\ \bibinfo {pages}
  {133401} (\bibinfo {year} {2012})}\BibitemShut {NoStop}%
\bibitem [{\citenamefont {D\"oppner}\ \emph {et~al.}(2010)\citenamefont
  {D\"oppner}, \citenamefont {M\"uller}, \citenamefont {Przystawik},
  \citenamefont {G\"ode}, \citenamefont {Tiggesb\"aumker}, \citenamefont
  {Meiwes-Broer}, \citenamefont {Varin}, \citenamefont {Ramunno}, \citenamefont
  {Brabec},\ and\ \citenamefont {Fennel}}]{DoePRL10}%
  \BibitemOpen
  \bibfield  {author} {\bibinfo {author} {\bibfnamefont {T.}~\bibnamefont
  {D\"oppner}}, \bibinfo {author} {\bibfnamefont {J.}~\bibnamefont {M\"uller}},
  \bibinfo {author} {\bibfnamefont {A.}~\bibnamefont {Przystawik}}, \bibinfo
  {author} {\bibfnamefont {S.}~\bibnamefont {G\"ode}}, \bibinfo {author}
  {\bibfnamefont {J.}~\bibnamefont {Tiggesb\"aumker}}, \bibinfo {author}
  {\bibfnamefont {K.-H.}\ \bibnamefont {Meiwes-Broer}}, \bibinfo {author}
  {\bibfnamefont {C.}~\bibnamefont {Varin}}, \bibinfo {author} {\bibfnamefont
  {L.}~\bibnamefont {Ramunno}}, \bibinfo {author} {\bibfnamefont
  {T.}~\bibnamefont {Brabec}}, \ and\ \bibinfo {author} {\bibfnamefont
  {T.}~\bibnamefont {Fennel}},\ }\href@noop {} {\bibfield  {journal} {\bibinfo
  {journal} {Phys. Rev. Lett.}\ }\textbf {\bibinfo {volume} {105}},\ \bibinfo
  {pages} {053401} (\bibinfo {year} {2010})}\BibitemShut {NoStop}%
\bibitem [{\citenamefont {Last}\ and\ \citenamefont
  {Jortner}(2000)}]{LasPRA00}%
  \BibitemOpen
  \bibfield  {author} {\bibinfo {author} {\bibfnamefont {I.}~\bibnamefont
  {Last}}\ and\ \bibinfo {author} {\bibfnamefont {J.}~\bibnamefont {Jortner}},\
  }\href@noop {} {\bibfield  {journal} {\bibinfo  {journal} {Phys. Rev. A}\
  }\textbf {\bibinfo {volume} {62}},\ \bibinfo {pages} {013201} (\bibinfo
  {year} {2000})}\BibitemShut {NoStop}%
\bibitem [{\citenamefont {Krishnamurthy}\ \emph {et~al.}(2006)\citenamefont
  {Krishnamurthy}, \citenamefont {Jha}, \citenamefont {Mathur}, \citenamefont
  {Jungreuthmayer}, \citenamefont {Ramunno}, \citenamefont {Zanghellini},\ and\
  \citenamefont {Brabec}}]{KriJPB06}%
  \BibitemOpen
  \bibfield  {author} {\bibinfo {author} {\bibfnamefont {M.}~\bibnamefont
  {Krishnamurthy}}, \bibinfo {author} {\bibfnamefont {J.}~\bibnamefont {Jha}},
  \bibinfo {author} {\bibfnamefont {D.}~\bibnamefont {Mathur}}, \bibinfo
  {author} {\bibfnamefont {C.}~\bibnamefont {Jungreuthmayer}}, \bibinfo
  {author} {\bibfnamefont {L.}~\bibnamefont {Ramunno}}, \bibinfo {author}
  {\bibfnamefont {J.}~\bibnamefont {Zanghellini}}, \ and\ \bibinfo {author}
  {\bibfnamefont {T.}~\bibnamefont {Brabec}},\ }\href@noop {} {\bibfield
  {journal} {\bibinfo  {journal} {J. Phys. B}\ }\textbf {\bibinfo {volume}
  {39}},\ \bibinfo {pages} {625} (\bibinfo {year} {2006})}\BibitemShut
  {NoStop}%
\bibitem [{\citenamefont {Mansbach}\ and\ \citenamefont
  {Keck}(1969)}]{ManPRev69}%
  \BibitemOpen
  \bibfield  {author} {\bibinfo {author} {\bibfnamefont {P.}~\bibnamefont
  {Mansbach}}\ and\ \bibinfo {author} {\bibfnamefont {J.}~\bibnamefont
  {Keck}},\ }\href {\doibase 10.1103/PhysRev.181.275} {\bibfield  {journal}
  {\bibinfo  {journal} {Phys. Rev.}\ }\textbf {\bibinfo {volume} {181}},\
  \bibinfo {pages} {275} (\bibinfo {year} {1969})}\BibitemShut {NoStop}%
\end{thebibliography}
\end{document}